\documentclass{ws-acs}
\usepackage{url}
\usepackage{longtable}
\usepackage[numbers]{natbib}
\usepackage{rotating}
\usepackage{epsfig}

\begin{document}
\title{The Dynamics of Internet Traffic: Self-Similarity, Self-Organization, and Complex Phenomena}
\author{Reginald D. Smith}
\address{PO Box 10051\\Rochester, NY 14610\\rsmith@bouchet-franklin.org}
\date{August 30, 2010}

\maketitle

\begin{abstract}
The Internet is the most complex system ever created in human
history. Therefore, its dynamics and traffic unsurprisingly take on
a rich variety of complex dynamics, self-organization, and other
phenomena that have been researched for years. This paper is a
review of the complex dynamics of Internet traffic. Departing from
normal treatises, we will take a view from both the network
engineering and physics perspectives showing the strengths and
weaknesses as well as insights of both. In addition, many less
covered phenomena such as traffic oscillations, large-scale effects
of worm traffic, and comparisons of the Internet and biological
models will be covered.
\end{abstract}


\section {Introduction}

In the last ten years, research of networks, especially the
Internet, has exploded amongst physicists. This follows at least
twenty years of research on the same and similar questions among
computer scientists and network engineering researchers. Interest in
the physics community began with several seminal papers on
small-world and scale-free networks \cite{faloutsos1, watts1,
Barabasi1, vespignani1, newman1}. Since then, this research has
progressed at a rapid pace borrowing well-developed tools from
statistical mechanics and thermodynamics, spectral graph theory, and
percolation theory among others to enhance the understandings of
fields such as Internet topology and social network analysis
previously only researched by the network engineering and sociology
communities. This new interdisciplinary work has been very fruitful.
However, there are gaps only more recently being addressed. In
particular, while the contribution of physicists to the
understandings of topology and community structure in networks is
substantial, often both physicists (and their counterparts in
sociology or network engineering) are often unaware of work on the similar questions in other fields. In addition, similar theoretical understandings and
predictions for network dynamics such as those now being reached for
topology remain elusive and are still in the earliest stages. There
have been good papers written on dynamics by physicists but
they have yet to formulate results with the same generality or detail
as the results on topology.

Just like the fist tentative measurements of Internet
topology allowed the field of networks to grow, Internet traffic
dynamics have provided a similar opportunity. Serious research on
the macroscopic nature of Internet traffic can be traced almost to
its inception, however, only in about the last 15 years has the
field come of age and begun to provide truly deep insights into how
communication over the largest technological edifice in human
history operates. Within this time the terms, ``self-similarity'',
``multifractal'', and ``critical phenomena'' have emerged to try to
refashion our ideas about the Internet and how it behaves.  This
paper could just be a review of the work by physicists and engineers on network traffic, however, my familiarity with the
research in the field has convinced me that for a full understanding
of the state of the art research in Internet traffic dynamics,
separating the views of engineers and physicists would make any
analysis incomplete and inadequate. An area for improvement in this
research on network traffic is the increasing collaboration and
cross-citation of works from other fields. Though in the study of
networks there are notable exceptions, in general, physicists and
engineers studying the Internet conduct their own research projects,
using field specific methodologies, and publish in field specific
journals with little cross-citation of relevant results from the
other disciplines. Indeed, one can see from the average paper in
physics journals such as Physical Review E or Physica A or from
engineering journals such as the IEEE or ACM series of journals that
many interrelated problems are being studied from totally different
perspectives. This has sometimes caused tension, especially between
physicists and network engineers, about the utility, details, or
real-world validation of physics based theories and the lack of
generality and generally applicable network principles among the
engineering perspective.

\begin{figure}[ht]
\centering
    \includegraphics[height=4in, width=5in]{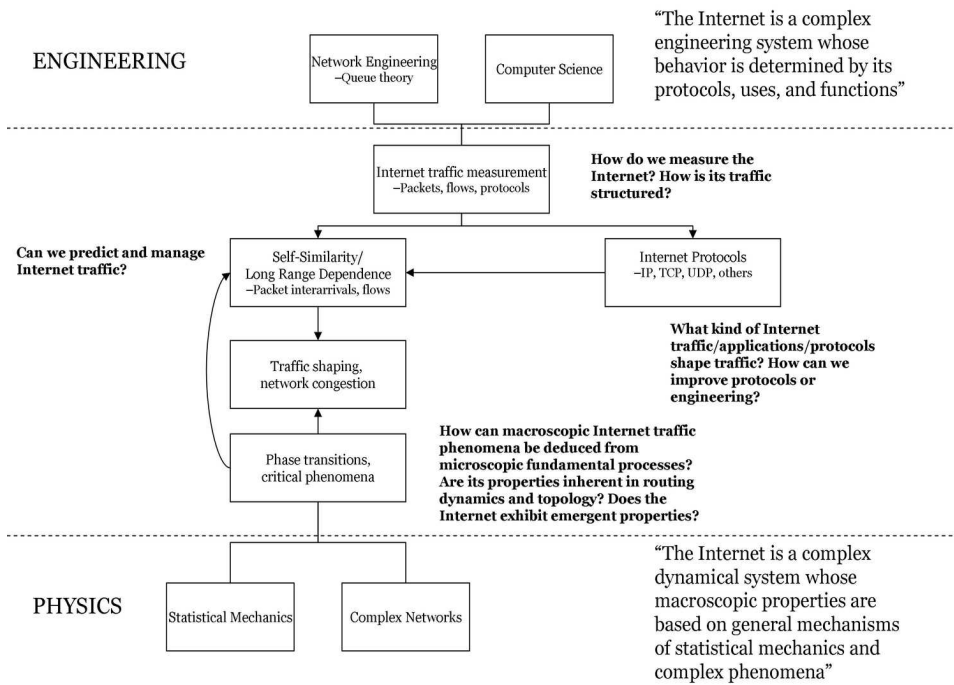}

        \caption{Network engineers and physicists often have diverging viewpoints on similar Internet traffic phenomena due largely
        to their backgrounds and training as well as the fundamental questions they ask. Here is a summary of those perspectives. On the top and bottom
        are the background knowledge and viewpoints of each side and in the middle are the problems they typically tackle and how they ask the questions.}
    \label{viewpoints}

\end{figure}

In figure \ref{viewpoints}, I have tried to give a full diagram and
summary of how these different world views operate. If someone is
trying to see which view is completely ``right'' or ``wrong'' they
are missing the point that the Internet is in some part everything
that both sides describe it as. It is an example of an engineering
system that is dependent on the precise nature of its protocols and other
workings to function. It is also a large-scale self-organized system
not far removed from those that physicists have studied in physical
systems for years. However, the research rarely reflects a full
synthesis of both views. Also, both perspectives are more correct on
some points than on others. For researchers interested in this area,
or even those who feel themselves to be seasoned veterans, one of
the first papers I recommend is \cite{caidameet} which is a short
position paper released by the 2006 CAIDA Workshop on Internet
Topology. Attended by eminent physicists and engineers, this
workshop clearly spelled out the promise and peril of
multidisciplinary research on Internet topology and traffic.

In line with the increasing focus on network dynamics, and the
reality that many of such research projects involve the Internet,
this review paper is meant to familiarize mainly physicists and
engineers with the major results of each other and how they
interrelate. For physicists, hopefully it will provide more exact
information on the workings of packet switching systems in the
Internet in order to allow us to better test our predictions against
reality, build more realistic models and simulations, validate
models and theories with real data such as traffic traces, and
contribute to the study of network dynamics through a more complete
understanding of the dynamics of the Internet. For network
engineers, they can see the issues raised by statistical mechanics
approaches towards network features such as congestion and realize
that there can possibly be large-scale phenomena quite independent
of detailed technical specifications. Since the Internet probably
has the largest readily accessible and easily understandable archive
of network traffic dynamics, it could likely play a huge role in
empirically validating theoretical ideas and simulations of dynamics
in networks.

This paper is organized in order to provide not only a review of
current research in the field but to provide a basic introduction in
the workings of the Internet for newcomers to the field. First, I
will review the basic ideas of Internet traffic including packets,
definitions of flows and throughput, and the basic protocols. While
this may seem common knowledge, much work in the field can only be
understood if you have the correct definitions and knowledge of the
Internet basics. Therefore, this is provided to prevent confusion
and perhaps inform on less discussed topics. Next, we will discuss
the evolution and composition of Internet traffic as far as usage
and protocols are concerned and study the basic dynamics of packet
flow including packet size distribution and Internet flow
characteristics. The meat of the paper involves a detailed
discussion of the self-similar nature of Internet traffic and how it
is defined and measured, the detailed workings and dynamics of the
TCP transport protocol, as well as the large body of work by
physicists on phase transition and critical phenomena models on
packet switching networks. These techniques will be reviewed and
their possible promises and current pitfalls addressed. Finally,
several interesting and related phenomena present in Internet
traffic such as oscillations will be covered. Each idea is given a
firm grounding and a thorough introduction but there will be no
pretense that I can completely delve into all research on any of
these ideas in one paper. Self-similar traffic alone has already
inspired several volumes on even its most esoteric aspects. However,
it is hoped this paper will allow someone with a reasonably
technical background and minimal familiarity with the subject and
research to quickly grasp the main themes and results that have
emerged from the research. Even for those that consider themselves
experts, there may be small insights or details that have been
poorly covered in most treatments and may add to their knowledge of
the subject.

\section{Packets, OSI Network Layers, and Key Terminology}
This first section of the paper will be its longest and will explain from the ground up the nature of Internet packet traffic and its key transport level protocol, TCP. This information is essential to those attempting to understand the Internet's functions and allows one to both understand other work in papers across many fields as well as develop more realistic models of the Internet based off its actual parameters and not idealizations.

\subsection{Packets and OSI}
In 1969, the Internet (then ARPANET) was first established as a
distributed packet communications network that would reliably
operate if some of its nodes were destroyed in an enemy attack as
well as facilitate communication between computer centers in
academia. Though the Internet has changed greatly up until today,
its packet switching mechanism and flexibility remain its key
aspects. The packet is the core unit of all Internet traffic. A
packet is a discrete bundle of data which is transmitted over the
Internet containing a source and destination address, routing
instructions, data description, a checksum, and data payload. Packet
handling and traffic management are governed by a complex set of
rules and algorithms collectively defined as a protocol.

Different protocols are responsible for handling different aspects
of traffic. Though this may seem trivial, protocols heavily affect
the nature of traffic and models of traffic which may be completely
valid for one protocol can be completely invalid for another. Also,
protocols are used in different applications or tasks and these
should inform any analysis of Internet traffic or predictive models
describing its behavior.

In addition, there are levels of tasks handled by certain protocols
and not others. These are broken out, traditionally into seven layers,
by a model known as the Open Systems Interconnection (OSI) model.
These seven layers are shown and described with examples in table 1.
For analysis of packet data, the application, transport, network,
and data link layer are typically the most relevant.

\begin{table*}[ht] \vspace{1.5ex}
\centering 
\scriptsize{
\begin{tabular}{|c| c| c| c|}
\hline
Layer&Number&Description&Example Protocols\\
\hline Application&7& Network applications such as terminal
emulation and
file transfer & HTTP, DNS, SMTP\\
 \hline Presentation&6&Formatting
of data and encryption &SSL\\
 \hline
 Session&5&Establishment and
maintenance of sessions & TCP sessions\\
\hline Transport&4&Provision of reliable and unreliable end-to-end
delivery &TCP, UDP\\
 \hline
 Network&3&Packet delivery, including
routing&IP\\
\hline Data Link&2&Framing of units of information and error
checking& Ethernet, ATM \\
\hline Physical&1&Transmission of bits on the physical
hardware&10BASE-T, SONET, DSL\\
 \hline
\end{tabular}
\label{OSItable} \caption{Breakdown of 7 layer OSI model.
Descriptions taken from \cite{IBM TCP}} \vspace{4.5ex} }
\end{table*}

The higher layers (higher number) always initiate a lower level
protocol. For example, for e-mail using the application protocol
SMTP, SMTP starts a TCP connection which itself uses IP packets to
deliver data.

Even within the same layers though, protocols can function much
differently. By far the most well-known and widely used suite is the
transport/network protocol combination TCP/IP. Transmission Control
Protocol(TCP), which manages sessions between two interconnected
computers, is a connection based protocol which means it has various
means of checking and guaranteeing delivery of all packets. This is
why it is widely used to transmit web pages using Hypertext Transfer
Protocol (HTTP), email with Simple Mail Transfer Protocol (SMTP),
and other widely used applications. TCP's connectionless cousin is
User Datagram Protocol (UDP). UDP sends packets without bothering to
confirm a connection or receipt of packets. This can make it
unreliable for delivery but much faster and more useful for
real-time applications like voice over IP (VoIP). TCP will be
covered in more detail and its differences elaborated later in the
paper. These differences cause TCP to react to feedback in its
traffic patterns and adjust its throughput based on these
considerations.

\subsection{Packet structure}

\begin{figure}[t]
\centering

    \includegraphics[height=0.5in, width=1.5in]{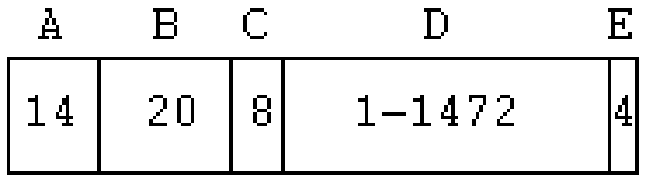}

        \caption{Structure of a packet in this paper. Proportions based on a 50 byte UDP packet payload. Numbers are size of headers or
        payload in bytes. A is the Link Layer (i.e. Ethernet) header which contains MAC address source and destination and payload type,
        B is the Internet Protocol (IP) header, C is the transport layer TCP/UDP protocol header, D is the
        data payload and E is the Link Layer (i.e. Ethernet) CRC checksum hash to prevent accidental corruption of the frame.}
\label{packetstructure}
\end{figure}

Packets have two main parts: a data payload which contains specific
data being transmitted and overhead which contains instructions
about packet destination, routing, etc. Each level and protocol has
a different amount of overhead as shown in table
\ref{packetstructure}. Overhead usually has both a fixed and
variable portion. However, for most transport and network layer
protocols the fixed portion can usually be considered the size of
the entire header. When dealing with the total size of packets and
measuring throughput, one must be careful to specify whether or not
the packet size includes overhead. Also, at the data link layer,
there is a maximum frame size of 1500 bytes in most systems (minus
data link layer overhead). The effect of packet size and packet size
distribution will also be covered in more detail later in the paper.

\begin{table}[h]
\centering 
\vspace{1.5ex}
\begin{tabular}{|c |c| }
\hline
Protocol&Header Size \\
\hline
IP&20 bytes for IPv4, 40 bytes for IPv6 \\
TCP&Normally 20 bytes, can be up to 60 bytes\\
UDP&8 bytes \\
Ethernet&14 bytes for header and 4 bytes for checksum\\
\hline
\end{tabular}
\label{packet} \caption{Packet header sizes for prominent protocols}
\vspace{1.5ex}
\end{table}

\subsection{Packet traffic characteristics}

Larger than individual packets is the packet flow which can be
statistically described using many important measures. Probably the
most widely known and important are bandwidth, throughput, goodput,
packet flow rate, flow, latency, packet loss, and Round Trip Time
(RTT).

Bandwidth - Bandwidth is the maximum possible throughput over a
link. Bandwidth, being an ideal, is almost never achieved under
normal conditions but provides a convenient benchmark to compare the
capacity of data links.

Throughput - Throughput is the rate of packet transmission over a
network link, usually in megabits per second (Mbps) or kilobits per
second (Kbps). It is the most widely recognized measure of network
data speed and essential in understanding the performance of data
traffic.

Goodput - Goodput is the measure of throughput excluding packet
overhead. When analyzing data, one must be careful to ascertain
whether traffic data is throughput or goodput. If it is goodput, the
total throughput is actually larger because you have to incorporate
the average packet overhead in the amount of data transferred.
However, for both throughput or goodput, the packet flow rate is the
same.

Packet Flow Rate - The rate of packet flow over a network link. This
differs from throughput or goodput in only measuring the number of
discrete packets that travel over the link, regardless of their
size. Throughput and packet flow rate are related by the following
equation

\begin{equation}
\label{basiceq} T = s\lambda
\end{equation}

Where s is the average packet size including overhead, T is the
average throughput, and $\lambda$ is the average packet flow rate.

Session - A TCP communication dialog set up between two computers by
first the delivery of a SYN packet, coordination request and
(SYN+ACK) packet, and finally an ACK by the initiating party.

Flow - This flow must be carefully distinguished from the packet
flow rate mentioned above. A flow in Internet traffic is defined
several ways but in general is a connection between a source and
destination which is continuously transmitting data. Usually, this
means a connection based protocol such as TCP where a connection is
made and data continuously transmitted until the connection times
out or a standard inter-packet arrival time is exceeded in the
``packet train'' \cite{packetflowdef1, packetflowdef2}. Sometimes
traffic instead of bytes is also measured in terms of the number of
flows. The distribution of flow sizes and their properties will be
covered later in the paper. Note when measuring flows using traffic auditing software such as Netflow or ARGUS, one must make sure to verify the exact method of flow measurement and make sure measurements across different systems are comparable and whether or not they match any of the theoretical definitions above.

Packet Loss - This is the percentage of all packets lost in transit.
It is usually measured as the percent difference between packets
transmitted in a packet flow and packets received on the other end
from the same packet flow. This affects all traffic and is caused by many causes that lead to the loss of packets including congestion, physical connection problems, and other issues. It has a large effect on TCP throughput.

Round Trip Time (RTT) - The statistical average time it takes a
packet to travel from a source to a destination and back. It is the
most common measure of latency on computer networks. It is closely
related to throughput and along with packet loss often used as a
measure of link congestion.

\subsection{Protocol traffic breakdown}

Most prominent studies of Internet traffic have tended to be done over high traffic transit networks or links linking smaller networks, such as those at a university, to the overall Internet. Overall, TCP dominates all traffic with about 95\% or more of total bytes, 85-95\% of all packets and 75-85\% of all flows using TCP as the transport protocol \cite{packetsize2,packetsize4,packetsize5}.
UDP comes second representing about 5\% or less of traffic with its
main function being sending DNS requests and communications. These results generally hold for the larger Internet, however, in measuring traffic protocol representation one must be careful to define whether a high traffic transit link, carrying traffic from many sources and applications, is used versus a link on an edge network closer to final users. Edge networks can have extremely particular traffic traces when there is large use of UDP for VoIP traffic or other multimedia type traffic.

TCP application traffic has generally evolved over time in three main
eras characterized by the dominant types of traffic influenced by
available applications and access speeds. In the Text Era
(1969-1994) most TCP traffic was driven by email, file transfers,
and USENET newsgroups. In 1989, C\'{a}ceres \cite{protocol1} at UC
Berkeley characterized Internet traffic of being 80\% TCP and 20\%
UDP by packet and 90\% TCP and 10\% UDP by bytes. TCP traffic bytes
were split roughly evenly between SMTP for email and FTP for file
transfer while UDP was mostly DNS. An updated study by C\'{a}ceres
and collaborators in 1991 \cite{protocol2} monitored traffic at
several universities again finding similar results. Once again at UC
Berkeley 83\% of packets were TCP, 16\% were UDP, and about 1\% were
ICMP. UDP traffic was predominantly for DNS at 63\% of its packets.
TCP traffic in terms of packets was 28\% telenet, 16\% rlogin (Unix
host login utility), 12\% FTP, 12\% SMTP, 12\% NNTP (USENET), with
the balance shared among other protocols. FTP was the largest
protocol in bytes at 36\% of all bytes. These dominant application
level protocols were confirmed by Claffy and Polyzos as well
\cite{protocol3}.

Next was the Graphics or Hyperlink Era (1994-early 2000). After CERN
made the World Wide Web free for any use in 1993, the graphics based
web grew rapidly. In 2004, Paxson \cite{protocol4} reported that in
Internet traffic though FTP, SMTP, and NNTP still held sway, HTTP
was by far the fastest growing protocol growing 300-fold in traffic
measured by connections in only two years and already vying to be
the 2nd most popular TCP application level protocol. By 1995, WWW
traffic had become the largest application level protocol with 21\%
of traffic by packets compared to 14\% for FTP, 8\% for NNTP and 6\%
for SMTP \cite{protocol4b}. By 1997, Thompson, Miller, \& Wilder
\cite{packetsize2} could report that HTTP dominated TCP traffic,
95\% of all Internet traffic bytes at this point, with 75\% of the
overall bytes, up to 70\% of the overall packets, and 75\% of the
overall flows during daytime hours. Its closest competitor, SMTP,
was reduced to only 5\% of packets and 5\% of all bytes. The
Internet was now a popular, mainly web and graphics based medium.

The current era is the Multimedia Era (early 2000-present). In this
period, sharing of multimedia through P2P file sharing applications
and streaming audio or video began to rival the web for dominance of
Internet traffic. A Sprint study on an IP backbone in early 2000
\cite{protocol5} reports that P2P was already rivaling the web in
terms of bytes transferred with at times P2P accounting for 80\% of
all traffic. Streaming also accounted for as much as 26\% of all
traffic as well. The web was still competitive, however, sometimes
accounting for 90\% of all traffic. By 2004, however, Fomenkov, et.
al. \cite{protocol6} could report WWW traffic clearly peaked in late
1999/early 2000 and P2P had dominated traffic growth ever since. A
recent April 2008 traffic trace study \cite{protocol7} shows the Web
and P2P sharing 34\% and 33\% of total TCP/IP bytes respectively.
However, P2P only accounts for about 3\% of all flows compared to
the 40\% of all flows dominated by the web showing P2P flows are
generally larger and more likely to be ``elephant flows''. Another
earlier study  by the same team \cite{protocol8} gave similar
results with Web and P2P (normal and encrypted) consisting of 41\%
and 38\% of bytes and 56\% and 4\% of flows respectively. 

One caveat on methodologies is necessary with more recent traffic. Though it is undisputable that multimedia and P2P type applications are carrying large amounts of traffic, identifying their exact nature based on port numbers is not always as exact as identifying HTTP traffic from port 80. There have been several methodologies created for traffic identification that use statistical measures in addition with TCP port numbers to determine traffic types. These include CoralReef (from CAIDA), BLINC\cite{ID1}, WEKA a suite of machine learning algorithms (from the University of Waikato, New Zealand), and graph based methods\cite{ID2}. Reviews of these methodologies evaluate their relative usage and show though port numbers are still the largest factor, they are often not completely reliable due to users or software intentionally or unintentionally changing port numbers for various hardware and applications \cite{ID3,ID4,ID5}.

\subsection{Topology}

As mentioned before, topology is currently the most studied feature
of the Internet and other computer networks by physicists. Due to
the wide range and depth of research being done in this field, this
paper will not present even a cursory review of its main ideas and
results. Dealing with Internet traffic, however, it is important to
balance the knowledgeable networking engineering perspective with
the more abstract methods used by physicists. For the now widely
familiar methods of physicists, the author recommends several
outstanding review papers \cite{review1, review2, review3, review4,
review5,review6,review7}. However, anyone measuring or analyzing
network topology network must read network engineering contributions
and rebuttals to many of the now standard scale-free network
techniques and theories. This is usually poorly covered or not at
all in the physics community. In particular, the author recommends
\cite{selfsimilar7, netcritic3,
netcriticnew1,netcriticnew2,netcriticnew2b, netcriticAMS, topologyIXP} which tackle the power-law distribution and ``robust yet fragile'' nature of the
Internet that physicists have theorized. A proposed alternative to general preferential attachment growth models for the Internet is given in \cite{HOTmodel,netcriticnew2}. Here the authors present a growth model that optimizes growth based on business and location considerations that match real Internet growth processes.

Also in \cite{netcriticnew3, netcriticnew5, netcriticnew6} the issues of
data integrity in measuring the Internet topology are described
including the errors introduced by several common methods such as
traceroute and BGP router tables. Essential reading for anyone
looking at this area. Finally, as a guide before claiming anything
as a power law, the author urges these two excellent articles on
tackling this determination in a rigorous fashion
\cite{powerlawsense1,powerlawsense2,powerlawsense3}.

\subsection{Packet Sizes}
\subsubsection{Distribution of packet sizes}
Internet traffic, with its various protocols and traffic types, has
many widely varying packet sizes. However, there is an upper limit
to packet size and this is almost always determined at the data link
layer. Various data link communications schemes, such as Ethernet or
ATM, impose an upper bound on the size of transmitted packets
through the hardware or operating system settings. This upper bound
packet size is often designated the Maximum Transmission Unit (MTU)
at the link layer.

In Ethernet, the current MTU on most systems is 1500 bytes. Packets
at the data link layer are often termed frames but the idea is the
same. This 1500 bytes includes the payloads and headers of all lower
level protocols but does not include the Ethernet frame header and
footer. Several studies on packet size distributions have shown that
packet size is in general a bimodal or trimodal distribution with
most packets being small (500 bytes or less)
\cite{packetsize1,packetsize2,packetsize3,packetsize4,packetsize5}.
In addition, the distribution of packet sizes is not a smooth
long-tailed distribution in that some packet sizes can predominate
due to system defaults.

For example, \cite{packetsize1, packetsize2} describe that there are
peaks in the frequency distribution for packet sizes. In a traces of
data over a  day or longer on a data link, they explain many reasons
for the small packet size. First, for TCP systems there is a
protocol option for ``MTU discovery'' that tries to find the MTU of
the network in order to make packets as large as possible. If MTU
discovery isn't implemented, TCP often defaults to an MTU of 552 or
576 bytes. Also, nearly half of the packets are 40-44 bytes in
length. These packets are used by TCP in control communications such
as SYN or ACK traffic to maintain the connection between the source
and destination systems. At 576 bytes \cite{packetsize1} the packet
size increases linearly to 1500 bytes showing that packet sizes in
the intermediate region are relatively equally distributed. In
general, according to \cite{packetsize2, packetsize4,packetsize5}
about 50\% of the packets are 40-44 bytes, 20\% are 552 or 576
bytes, and 15\% are 1500 bytes. Table \ref{packetsizetable} shows
the distributions of packet sizes from a traffic trace and fit well
with the studies except the absence of a strong peak in the 552 or
576 byte range.

\begin{table}[ht]
\centering 
\begin{tabular}{c c c c }
\hline
Packet Size Range&ALL&TCP&UDP\\
\hline
0-19&0\%&0\%&0\%\\
20-39&2\%&0\%&0\%\\
40-79&59\%&69\%&19\%\\
80-159&7\%&2\%&23\%\\
160-319&3\%&1\%&15\%\\
320-639&7\%&3\%&34\%\\
640-1279&3\%&3\%&6\%\\
1280-2559&18\%&22\%&4\%\\
\hline
\end{tabular}
\label{packetsizetable} \caption{Packet size distribution of a
capture of 1 million packets in a 100 second trace from the MAWI
\cite{mawi} traffic trace archive from Samplepoint-B on July 22,
2005.}
\end{table}

Kushida \cite{packetsize5}, despite being one of the only papers that looks at
packet size distribution among UDP separately, clarifies that
since 98.2\% of the traffic measured in the paper is TCP, the UDP
contribution to overall IP and Internet traffic packet size
distribution should be considered negligible. Using a different
measurement for packet size distribution, that looks at the ratio of
packet size*number of packets versus the total traffic measured,
Kushida finds a series of peaks between 75 and 81 bytes and another
large peak at 740 bytes. None of these peaks are substantial,
however, and no size of packet reaches even 10\% of the total. Since
UDP has no connection based features such as TCP, the reason for
these peaks is not inherent in the protocol itself. UDP is mainly
used for Domain Name Server (DNS) and Simple Network Management
Protocol (SNMP; a network monitoring protocol) and applications
related to these functions drive the size of the UDP packets.

Finally, there is often an asymmetry in packet size for both
directions in a flow. For example, if a web page is being served to
a PC, the PC will be receiving large TCP packets with HTTP (WWW)
data while it will only be sending comparatively smaller packets as
data requests back. Packet size also has diurnal variations where it
can be larger during daytime hours and in international links,
\cite{packetsize1} showed that the average packet size on both
directions of the link oscillated out of phase by about 11 hours
(2.9 radians).

Does packet size or MTU matter? Absolutely, in fact many network
engineers realize that average packet size and MTU are critical
factors in determining the overall maximum throughput in a network.
Recalling equation \ref{basiceq}, for a fixed throughput, decreasing
the packet size increases the packet flow rate. Oftentimes, many
believe that the key throttle in computer network throughput is its
stated bandwidth. In fact, bandwidth bottlenecks are rarely the
bottleneck on network performance. Computer network hardware
typically has a maximum packet flow rate it can effectively handle,
afterwards packets begin forming queues in the hardware buffer and
congestion reduces throughput. Smith, in \cite{transcritical} showed
how on a normal Ethernet link between two computers, the maximum
throughput across varying packet sizes exhibited a transcritical
bifurcation.

In fact, one problem currently plaguing next generation high speed
networks is outdated, smaller MTUs on the systems of their users.
Therefore, in order to take advantage of the increasing bandwidth
capabilities of the Internet, there is a concerted push in some
corners to raised the typical MTU above the normal Ethernet 1500
bytes, up to 9000 bytes where possible, to allow more rapid
communication. There will need to be larger studies on network
hardware such as routers to understand how the MTU completely
affects throughput and whether the bifurcating behavior is present
for saturated links in the Internet at large.

\subsection{Flow Size Structure and Distribution}

\subsubsection{Definition and nature of flows}
As mentioned in the definition of an Internet flow, a flow is
defined as continuous communication between a source and destination
system. Flows are typically described by one of two definitions: an
identifiable clustering of packets arriving at a link or by
identifying characteristics such as the source and destination
addresses along with an identifying label such as a TCP session ID
or an IPv6 flow label.

For the first definition, the most widely used definition was given
by Jain and Routher in 1986 \cite{packetflowdef1} while studying
data on a token ring network at MIT. While also noting that the
interpacket arrival rate is neither a Poisson or compound Poisson
distribution, they defined individual flows as ``packet trains''
where a packet train is defined as a sequence of packets whose
interarrival times are all less than a chosen maximum interarrival
gap, usually determined by system software and hardware
configurations. If a packet is received after a longer interval than
the maximum gap, it is considered part of a new flow. This brings up
one important characteristic of Internet flows: though they
obviously have a time average, they are extremely bursty and
inhomogeneous compared to most other types of flows studied in
physics.

For the second method, the first and still likely most widely used
method of identifying flows via address or label is using TCP
packets. TCP flows start with a SYN packet and end with a FIN
packet. Therefore matching SYN and FIN packets with source and
destination IP addresses and session ID in the TCP headers are often
used to define flows. Another elaborate definition was presented by
Claffy et. al. \cite{packetflowdef2} who represent a flow as active
if there is interpacket time less than a maximum value and
distinguish flows by a group of packets identified by aspects
including source/destination pairs, unidirectional nature (flows in
only one direction), protocols used, and other factors that may
distinguish the packet destinations.

The new next generation Internet Protocol, IPv6, though not yet
implemented widely beyond a now defunct test network called 6Bone,
has been designed with a part of its header overhead reserved for a
``flow label''. This flow label would allow the traffic source to
provide a unique identifier that would clearly distinguish IP
traffic flows. Besides improvements in routing and traffic
management, this will allow more accurate research as IPv6 is
implemented throughout the Internet.

\subsubsection{Distributions of flow characteristics}

Early in the paper, it was mentioned that long-tail behavior is
present in Internet traffic to the same extent as it is in the
topology. Flows are no exception and several quantities used to
describe flows have long-tail distributions. In particular, the
distributions of sizes in terms of data transferred, duration in
terms of length of the flow, and data rate of flows have all been
found to exhibit long-tail distributions.

These flows have been given certain names throughout the literature
which are summarized by Lan and Heidemann \cite{elephant0}. Flow
sizes are divided into two classes: ``elephants'' and ``mice ``where
elephants  are a small part of all flows measured over a certain
time but account for a large number of the bytes transferred while
many other flows account for proportionally smaller components of
the overall traffic (the mice). Elephant flows have been described
in detail in several papers \cite{elephant0,elephant1, elephant2,
elephant3, elephant4, elephant5}, in particular in a paper by Mori
et. al. describes a traffic trace where elephant flows are only
4.7\% of all flows measured but 41\% of all traffic during the
period. Barth\'{e}lemy et. al., \cite{elephant6} give a related
result studying routers on the French Renater scientific network.
They conclusively find that a small number of routers (a so-called
'spanning network') submit the vast majority of data on the network
while the contribution of the other routers is exponentially
smaller.

Elephant flows, though agreed upon in principle, have been defined
differently in many papers. Estan \cite{elephant1} defined an
elephant flow as a flow that accounts for at least 1\% of total
traffic in a time period. Papagiannaki \cite{elephant5} uses flow
duration to classify elephant flows. Lan and Heidemann
\cite{elephant0, elephant3} use a statistical definition where a
flow is considered an elephant flow if the amount of data it
transmits is at least equal to the mean plus three standard
deviations of flow size during a period. This is 152kB in their
paper. This final definition implicit assumes the scaling exponent
among flow sizes, $\alpha$ is at least 2, since the variance for the
distribution is infinite if $\alpha < 2$.

\begin{figure}[ht]
\centering
    \includegraphics[height=3.5in, width=3.5in]{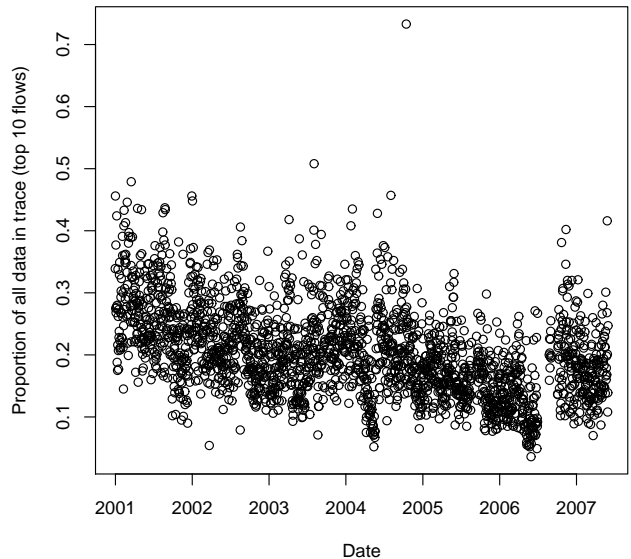}
\caption{Percent of data in all flows occupied by the top 10 flows
over time. From the WIDE MAWI traffic trace archive \cite{mawi}
using data from Samplepoint-B from 12/31/2000 - 5/31/2007. Median
daily flows total about 350,000} \label{elephantgraph}
\end{figure}

In figure \ref{elephantgraph}, the author has used data from the
WIDE MAWI \cite{mawi} traffic trace archive which measured the daily
traffic over a T-1 line between Japan and the Western US to show the
relative proportion of all traffic the top 10 flows represented over
time from 2001-2007. The upward tick in mid-2006 reflects the
upgrading of the data link speed from 100Mbps to 1Gbps. The \% of
all traffic captured by the top 10 flows declines over time as the
number of overall flows per day increases and the top 10 occupy a
declining share of the number of flows.

Research from Mori et. al. \cite{elephant4} also gives evidence that
elephant flows not only occupy disproportionate amounts of traffic
but they are also more likely than mice to be responsible for
congestion in links.

The duration of flows has been classified with similar zoological
flair. Most flows have a relatively short duration while a small
number of flows have a comparatively very long duration. Brownlee
and Claffy \cite{dragonfly} analyzed duration among Internet
streams, which are individual IP sessions versus one way flows of
packets typifying flows. About 45\% of streams were of a very short
duration, less than 2 seconds, and were termed ``dragonflies''.
Short streams were defined has having a duration from 2s to 15
minutes and consisted of another 53\% of all flows. ``Tortoises''
were flows with a duration greater than 15 minutes and accounted for
1-2\% of all streams but 50\% of all bytes transferred. The
dragonfly/tortoise definition is simplified and extended to flows in
\cite{elephant0} where a dragonfly is a flow less than the mean flow
duration plus three standard deviations which is 12 minutes in their
paper. They find 70\% of all Internet flows are less than 10
seconds.

Lan and Heidemann \cite{elephant0} also introduce a new measure of
flow, ``cheetahs'' and ``snails'' to characterize the distribution
of throughput in flows. Cheetahs are flows with an average
throughput greater than the mean plus three standard deviations.
Their dividing throughput is 101 kB/s in the paper. According to
their measurements, about 80\% of Internet flows have a throughput
of less than 10 kB/s.

These different types of measurements on flows are obviously not
independent and in fact are heavily correlated in several ways.
Cheetahs tend to be high throughput but small in size and short in
duration Zhang \cite{elephant6} previously showed a correlation
between flow size and rate and Lan and Heidemann \cite{elephant0}
confirm this showing 95\% of cheetah flows are dragonflies with a
duration of less than 1 second. 70\% of cheetah flows are also
smaller than 10 kB. Elephant flows tend to be large in size and
duration but low in throughput. Only 30\% of elephant flows in
\cite{elephant0} are faster than 10 kB/s and 5\% are faster than 100
kB/s. 50\% of elephant flows lasted longer than 2 minutes and 20\%
of elephant flows lasted at least 15 minutes.

Different flow types are also dominated by different types of
traffic. Elephants seem to be mostly web and P2P traffic while Tortoises
are mostly DNS. Cheetahs have by far mostly web and DNS traffic. This leads to the question as to the causes of bursty flows (as differentiated from bursty overall traffic described in the section on self-similarity). TCP mechanisms, described in the next section, could cause the burstiness of some flows, however, it could also be driven by the bursty data requests from user applications such as interactive applications like MMORPGs.

The granularity and nuance in the characteristics of flows is an
interesting theoretical and practical challenge for those studying
Internet dynamics. But it still gets even better as our next section
on the dynamics of TCP traffic demonstrates.

\begin{table}[ht]
\centering 
\begin{tabular}{c c c c c }
\hline
Category&Large-size&Long-lived&Fast&Bursty\\
\hline
Elephant&Y&Y&N&N\\
Tortoise&N&Y&N&N\\
Cheetah&N&N&Y&Y\\
\hline
\end{tabular}
\label{table:animalflows} \caption{Classification and description of
flows from Lan and Heidemann \cite{elephant0}}
\end{table}

\section{Explanation and Mechanics of TCP}

Now that we understand the basics of packet traffic, we can dig more deeply into one of the key functions of the Internet: connection based packet delivery through TCP. As stated earlier, TCP by far is the bulk of Internet traffic.
Therefore, any discussion on Internet traffic is by and large a
discussion of TCP/IP traffic. TCP is a connection based protocol and
relies on several programmed algorithms to manage and guarantee the
delivery of packet traffic. The fact is, however, TCP was developed
when the Internet was relatively small. Though it is still useful
and efficient, the large-scale macroscopic effects of its operation
were not easily predictable and were only measured or derived later.

Volumes of articles have been written on TCP behavior, possible
algorithmic improvements, and traffic management. TCP has several
features including buffering and congestion control that allow it to
be one of the only Internet protocols that uses feedback to adjust
protocol performance. Nonlinear effects combined with feedback have
been well-known to produce complex systems phenomena and TCP is no
exception.  In this section, TCP's basic mechanisms will be defined
and explained and then linked with the most common theories of
network performance and congestion.

\subsection{A short explanation of TCP}

There are many good guides on TCP, but most information in this
paper is taken from an IBM guide \cite{IBM TCP}. TCP relies on
several key features which are necessary to ensure reliable and
smooth delivery of packets between the source and receiver. The TCP
connection starts out with a ``three-way handshake'' which consists
of one SYN (synchronize) and two ACK (acknowledge) packets. TCP
flows are based on the concepts of windows and flow control. When
packets are transmitted they are given sequence numbers to determine
the correct order of data transmission. The source then waits to
receive and ACK packets before transmitting additional packets.

The number of packets a source can transmit before needing to
receive at least one ACK packet is the window. When a TCP connection
is initiated and as it continues, the receiver sends an ACK packet
which lets the source know the highest sequence number it is able to
accept given buffer memory and system constraints. The source then
sends the number of packets to fit that window and waits for an ACK.
For every packet confirmed, an additional one is sent and the window
size is maintained. The window size can be changed by the receiver
in every ACK packet by varying the highest sequence number it can
receive so the window often varies over the course of a
transmission. If an ACK for a packet is not received within a
timeout period, TCP considers the unacknowledged packets lost and
retransmits them.

Because of the possible need for retransmission, TCP must buffer all
data that has been sent but has not received an ACK. The size of
this buffer at the sender is usually calculated by the bandwidth
delay product which is the product of the link bandwidth and the
RTT. Therefore, for high bandwidth links or long RTT links, the
buffer can become increasingly large and burdensome on the operating
system.

\subsection{Congestion control}
\begin{figure}[ht]
\centering
    \includegraphics[height=3.5in, width=3.5in]{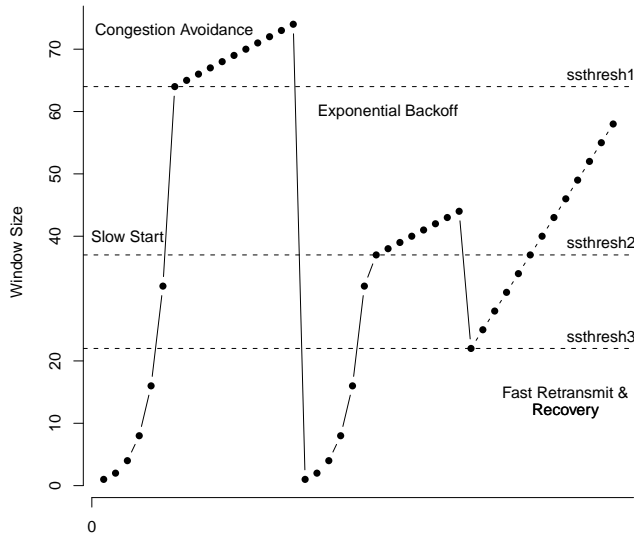}
\caption{Illustration of the window size of a hypothetical TCP flow over time. Each type of congestion control mechanism is illustrated in the figure. ssthresh1, ssthresh2, and ssthresh3 are the first, second, and third slow start thresholds over time respectively. ssthresh2 is 1/2 of the value where the timeout congestion occurred and ssthresh3 is 1/2 the value where the duplicate ACK congestion occurred. The second congestion is caused by duplicate ACKs and initiates fast retransmit and fast recovery.} \label{tcpgraph}
\end{figure}

Congestion control is one of the most prominent and differentiating
aspects of TCP as compared to UDP and any other transport level
protocol. Flow control uses ACK feedback to coordinate smooth
transmission between the sender and receiver and congestion control
uses feedback from the network throughput environment to adjust the
sender rate in order to not cause network congestion. Thus Internet
traffic, which is largely TCP/IP, behaves in part as a massive closed
loop feedback system between the transmission throughputs of
multiple senders which are modified by the measured traffic
congestion environment. This has doubtlessly led to much research in
self-organization in TCP traffic which will be discussed later.

The congestion control algorithm is not uniform across all TCP
software implementations and has various flavors named after resort
cities in Nevada including Vegas, Reno, and Tahoe. All
implementations though essentially share four general features: slow
start, congestion avoidance, fast recovery (not in Tahoe), and fast
retransmit.

Slow start is used to address the inherent problem that regardless
of what TCP window the receiver advertises in its ACK packets, the
network still may be so slow or congested in points as to not be
able to handle that many packets transmitted over such a short time.
Slow start handles this by overlaying another window over the TCP
window called the congestion window (cwnd). At first, this window
starts at one packet and tests to see if an ACK is received. If so,
the congestion window grows to two, waits for two ACKs, and then
grows to 4, increasing by powers of two at each successful step. This occurs until the window is the maximum size advertised by the receiver. Afterwards, the receiver's window size can increase by one segment per ACK successfully received. The sender will eventually use the smaller of the congestion window or
the window advertised by the receiver as the TCP window. This
insures that if the receiver's window is too large for current
network performance, the congestion window will compensate. Slow
start has a threshold though which is the maximum size possible for
any congestion window.

Congestion avoidance works in tandem with slow start. Congestion
avoidance assumes that any packet loss (packet loss is normally
assumed to be much less than 1\%) signals network congestion. As stated earlier, this is not necessarily true as hardware or network issues which cause lost or dropped packets and aggressive ``crowding out'' by large flows of aggressive connectionless protocols like UDP can also generate this effect. The
connection measures packet loss by a timeout or duplicate ACK
packets. If congestion is detected, congestion avoidance slows down
the TCP connection by setting the slow start threshold to one half
of the current congestion window, the so-called exponential backoff.
If a timeout caused the congestion, the congestion window is
``reset'' down to one packet and slow start repeats. Slow start
increases the window size to the new slow start threshold and then
congestion avoidance takes control of the congestion window size.
Instead of increasing the window size in an exponential manner,
congestion avoidance increases it in increments with every ACK
received according to the following equation

\begin{equation}
\frac{\mbox{segment size}}{\mbox{congestion window size}}*
\mbox{segment size}
\end{equation}

where the segment size is the size in bytes of data TCP fits into
each packet. Therefore the congestion window increases linearly
controlled by the congestion window algorithm.

Fast retransmit is where TCP uses the number of duplicate ACKs
received to determine whether a packet was received out of order at
the receiver or likely dropped. If three or more duplicate ACKs are
received it assumes the packet was lost and retransmits it. This
prevents TCP from having to wait the entire timeout period before
retransmitting. Since fast retransmit is based on the assumption of
a lost packet, congestion avoidance comes into play. However, fast
recovery takes over in the situation where fast retransmit is used
and allows the TCP window to not decrease all the way to one and
restart using slow start but by setting the threshold to one half of
the congestion window size and starting the congestion window size
at the threshold size + 3 x segment size. The congestion window then
increases by one segment for each additional ACK.

\subsection{TCP macroscopic behavior}

The intricacies of the operation of TCP have lead to much research
characterizing the protocol's average or expected performance and
its influence on the overall traffic patterns of the Internet. One
of the best-known and widely reported results is a famous equation
for the maximum possible throughput for a TCP connection developed
in early versions by Floyd \cite{TCPthrough1} and Lankshman \&
Madhow \cite{TCPthrough2} and in a most widely known version by
Mathis, Semke, and Madhavi \cite{TCPthrough3}. They explore the
expected performance of TCP against a background of random, but
constant probability, packet loss given the window resizing by the
congestion avoidance backoff mechanism. Their famous result (often
called the SQRT model) is for the theoretical maximum throughput for
a TCP connection and is given by

\begin{equation}
\label{TCPequation} T \leq \frac{\mbox{MSS}}{\mbox{RTT}}
\frac{C}{\sqrt p}
\end{equation}

where $MSS$ is the maximum segment size, typically defined by the
operating system for TCP and is usually 1460 bytes \cite{PingER},
$p$ is packet loss percentage, and $C$ is a constant which varies
based on assumptions of periodic or random period loss and the
handling of ACK by a congestion avoidance algorithm. Since C is
usually less than 1 the equation can be simplified to

\begin{equation}
\label{TCPequationsimple} T < \frac{\mbox{MSS}}{\mbox{RTT}}
\frac{1}{\sqrt p}
\end{equation}

This equation assumes packet loss is handled by congestion avoidance
and determined by receiving duplicate ACK packets, not packet loss
via timeouts. Though this is the most famous and widely used
equation, a more accurate one, especially when $p > 0.02$  was
introduced by Padhye, Firoiu, Towsley, and Kurose
\cite{TCPthrough4}. This equation, based on a version of TCP Reno,
also incorporates packet loss due to timeouts which is more
realistic for higher packet loss situations. Their equation yields
an approximation for throughput where

\begin{equation}
\label{TCPequationPadhye} T \approx \mbox{min}
\left(\frac{W_{m}}{\mbox{RTT}} \mbox{,}
 \frac{1}{\mbox{RTT}
\sqrt{\frac{2bp}{3}} + t_{0} \mbox{min}\left(1, 3
\sqrt{\frac{3bp}{8}}\right)p \left(1+32p^2\right)}\right)
\end{equation}

This equation is also known as the PFTK equation. Here $p$ is once
again packet loss, $W_{m}$ is the maximum window size advertised by
the receiver, $b$ is the number of packets acknowledged by each ACK
(usually 2), and $t_{0}$ is the initial timeout value.

In \cite{TCPthrough4} the authors compared the fit to their equation
versus the SQRT equation for real data and state PFTK fits better. A
more complete analysis and comparison was conducted by El Khayat,
Geurts, and Leduc \cite{TCPthrough5}. They note both equations
neglect slow start which makes them inappropriate for very short TCP
flows and the equations also neglect fast recovery. To test the
models they generated thousands of random networks with random graph
topologies where the number of nodes (10 - 600) was chosen at random
and the bandwidth (56 kbps - 100 Mbps) and the delay (0.1ms -500ms)
were chosen randomly for each link. They then tested the TCP
throughput on these virtual networks versus both equations and made
comparisons using the mean squared error, $R^{2}$, the over/under
estimation ratio of average calculated throughput to actual
throughput, and an absolute ratio which takes the larger of the
first ratio or its inverse. For all metrics, the PFTK equation
performed better but was still a poor predictor of actual TCP
performance giving incorrect estimates roughly 70\% of the time.

In addition to measuring the current macroscopic behavior of TCP, there are many studies attempting to model the behavior of TCP flows under alternate congestion management paradigms. The most prominent of these is TCP/RED (Random Early Detection), first proposed by Sally Floyd \cite{TCPred}. TCP/RED actively manages the router queue by increasing the probability of a router deliberately dropping a packet as the queue buffer becomes increasingly full in order to better manage congestion. There are several papers \cite{TCPred2,TCPred3,TCPred4} that state large-scale traffic patterns such as chaos or oscillations could be induced but these are still mostly based on theory or simulation.

\section{Packet Arrival Times - Self-Similarity, Long Range Dependence, and Multifractals}

\subsection{Self-similar traffic and long range dependence}
One of the most widely researched and discussed characteristics of
Internet data traffic among both the computer science and physics
communities is the self-similar nature of Internet packet arrival
times. Interestingly enough, the trajectory of research on this
begins with a shattering of simplistic preconceptions about network
traffic similar to that of Faloutsos and Barab\'{a}si and Albert regarding
Internet topology random graph models. Like pre-Barab\'{a}si/Albert theory assumed all telecommunications networks, including the Internet, were random
graphs, early research in Internet traffic regarded packet arrival
times as based on a Poisson (Erlang-1) or Erlang-k distribution
similar to that in telephone switching and call center traffic
\cite{traffichistory}. The first cracks in this were a paper by
Leland and Wilson \cite{selfsimilar1} which showed packet
interarrival times that seemed to exhibit both diurnal fluctuations
and did not seem to adhere to a Poisson distribution. The second
paper by Leland, Wilson, Taqqu, and Willinger \cite{selfsimilar2}
thoroughly and convincingly debunked the theory of Poisson arrival
of packets in Internet traffic and using rigorous statistics showed
that Internet traffic had self-similar characteristics and
correlations over long time scales (long-range dependence or LRD).
Like degree distributions in scale-free topologies, the packet
arrival per unit time exhibited long-tail distributions where large
bursts of traffic were not isolated and extremely rare statistical
coincidences but par for the course over all time scales. The
studies were based on four captures of data over four years. The
traffic traces, taken at the former AT\&T Bellcore research
facility, varied from 20 to 48 hours in length and recorded the
timestamps of hundreds of millions of packets.

The authors are also the first to describe Internet traffic as
having a fractal character. This research has since been confirmed
in a torrent of papers which are too numerous to describe. Paxson
and Floyd \cite{selfsimilar3} confirmed the failure of Poisson
modeling in long-tail traffic behavior in Wide Area Network (WAN)
data in several protocols including TCP, FTP, and Telnet. Crovella
and Bestavros \cite{selfsimilarcause3b} demonstrated long-tail
distributions in WWW traffic including packet interarrival times,
file download size distributions, file download transmission time
distributions, and URL request interarrival times. Other papers have
essentially confirmed in most cases that long-range dependence is a
key feature of Internet packet traffic.

\subsection{Measuring self-similarity and long range dependence}

There are several good review articles detailing the mathematical
techniques used to investigate self-similar processes in network
traffic \cite{selfsimilar5, selfsimilar6, selfsimilar7,
selfsimilar8, selfsimilar9, selfsimilar10, selfsimilar11,
traffichistory}. Here we will cover the most prevalent and important
ones.

The simplest definition for self-similarity assumes that for a
continuous time process, $X(t) \mbox{   }   t\geq 0$, for scaling the
time by a factor $c_1$,

\begin{equation}
X(t) = c_1^{-H}X(ct)
\end{equation}

where $H$ is the Hurst exponent and takes a value between 0 and 1
for self-similar processes. For a self-similar process that exhibits
long-range dependence, $H$ is between 1/2  and 1. This definition,
like most other for self-similarity, implicitly assumes a stationary
process.

The most accepted and widely used definitions are termed the
so-called first-order and second-order similarity. First order
similarity is based on the autocorrelation of the traffic trace.
Assuming a traffic trace is defined as a stationary stochastic
process $X$ with a set of values at time steps t:

\begin{equation}
X = (X_{t}: t = 0, 1, 2...)
\end{equation}

and the autocorrelation function, $\rho(k)$, is defined as

\begin{equation}
\rho(k) = \frac{E\left[(X_{t} - \mu)(X_{t+k} -
\mu)\right]}{\sigma^{2}}
\end{equation}

Where $\mu$ is the mean and $\sigma^{2}$ is the variance of the
traffic. The self-similar behavior is manifested in that the
behavior of the autocorrelation function is not one which
exponentially decays with time as with a short range dependent time
series but rather exhibits a power law behavior

\begin{equation}
\rho(k) \sim c_2k^{-\beta}   \mbox{   }        0 < \beta < 1
\end{equation}

Where $c_2$ is a positive constant and the approximation symbol
indicates this behavior is the asymptotic behavior of the system as
$k \rightarrow \infty$. Fitting a linear regression to an
autocorrelation or autocovariance plot should not be considered a
rigorous or best practice method of determining self-similarity and
the Hurst exponent. There are various other tools, with shortfalls
as well, that are best used to make an accurate determination.

\emph{Second order similarity/aggregated variance analysis}

Second order similarity is defined as taking the original time
series and recreating it for different time ``windows'' $m$ where
all time values in the series in windows of length $m$ are averaged.
For example, the new time steps become $t=0, m, 2m..,N/m$. Second
order similarity, also known as aggregated variance analysis, is
formally defined as taking the new time series

\begin{equation}
X^{(m)}_{k} = 1/m(X_{km -m+1}+\ldots+X_{km})
\end{equation}

for all $m = 1,2,3,\ldots$. The time series is called exactly
self-similar if the variance of $Var(X^{(m)}) =
\frac{\sigma^2}{m^{-\beta}}$ and

\begin{equation}
\rho^{(m)}(k) = \rho(k) \mbox{   }   k\geq0
 \end{equation}

For a normal independent and identically distributed time series the
variance would behave as $ Var(X^{(m)}) =\frac{\sigma^2}{m}$. With
self-similarity it decays much more slowly given the range of $ 0 <
\beta < 1$ . The time series is called asymptotically self-similar
if the autocorrelation function of the new time series for large k
behaves as

\begin{equation}
\rho^{(m)}(k) \rightarrow \rho(k) \mbox{   } m\rightarrow \infty
\end{equation}

For both definitions of self-similarity, the Hurst exponent $H$ can
be derived from the value of $\beta$ according to the equation $H =
1-\beta/2$. This confines the Hurst exponent to values of between
1/2 and 1 for a self-similar system. Note $H$ = 1/2  exponent is
identical to that of random Brownian motion and $H$ = 1 reflects
complete self-similarity.  In most studies, $H$ is estimated to be
around 0.8 in most types of Internet traffic. The results from the
data trace analyzed by the author in figure \ref{traffic} give a
Hurst exponent of 0.81.

One must take care to differentiate two similar but not identical
aspects of Internet traffic: self-similarity, just defined above,
and long-range dependence. Long-range dependence is defined as a
system where the autocorrelation function behaves as

\begin{equation}
\sum_{k}|\rho(k)| = \infty \label{LRD}
\end{equation}

when $H > 1/2 $ for self-similar traffic long-range dependence is
implied but in other conditions you can have long-range dependence
but not self-similarity as long as equation \ref{LRD} is satisfied.
Long-range dependence is also called persistence and is contrasted
by short-range dependence (SRD) which manifests in processes where
$0 < H < 1/2$. LRD also depends on an assumption of stationarity in
traffic which is reasonable on timescales of minutes to hours but is
less useful over large timescales due to diurnal traffic variations
and long-term trends.

\emph{R/S Statistic}

Again, we separate the time series into $m$ equal blocks of length
$N/m$ except all values in each block are aggregated by simple
summation. Define $n$ as $n = N/m$ and define the range $R(n)$ as
the difference between the value of the largest block and the
smallest block. Define $S(n)$ as the standard deviation of the
values of the blocks. The ratio $R(n)/S(n)$ should scale with $n$
such that

\begin{equation}E[R(n)/S(n)] \sim c_3n^H \label{R/S}
\end{equation}

Note that one problem with both the R/S and other methods such as
aggregated variance is choosing the right range for the sizes of the
blocks \cite{hurstguide}. Choosing values of $m$ that are too small
makes short term correlations dominate, while a large $m$ has fewer
blocks and gives a less accurate estimate of $H$. One approach
created to deal with this issue is wavelet analysis of the logscale
diagram which is covered in the next section on wavelet methods.

\emph{Periodogram}

An additional test for long-range dependence is the presence of
$1/f$ noise in the spectral density of the time signal at low
frequencies. The exponent of $1/f$ noise is related to $\beta$ as
well where $f(\lambda) = c\lambda^{-\gamma}$ where $c$ is a constant
(unrelated to previous ones), $\lambda$ is the frequency and $0
<\gamma < 1$ and $\gamma = 1- \beta$.

Often, the spectral density, $I(\lambda)$ is estimated as

\begin{equation}
I(\lambda) =  \frac{1}{2\pi
N}\left|\sum^N_{j=1}X_fe^{ij\lambda}\right|^2
\end{equation}

Whose log-log plot slope should be close to $1-2H$ near the origin.

\emph{Scaling of Moments}

In \cite{multifractal1}, the authors use the fact that the moments scale with the
length of the time series to identify self-similarity. Define the
absolute moment as

\begin{equation}
\mu^{(m)}(q) = E|X^{(m)}|^{q} =
E\left|\frac{1}{m}\sum^{m}_{i=1}X(i)\right|^{q}
\end{equation}

The absolute moment $\mu^{(m)}(q)$ scales as

\begin{equation}
\label{fracscale} \log{\mu^{(m)}(q)} = \beta(q)\log{m} + C(q)
\end{equation}

Where $\beta(q) = q(H-1)$.

\begin{table}[ht]
\centering 
\caption{Relationship among key exponents} \centering
\begin{tabular}{c }
\hline
$H = 1-\beta/2$\\
$\beta = 2(1-H)$\\
$\gamma = 1 - \beta$\\
$\beta = 1 - \gamma$\\
$\gamma = 2H-1$\\
$H = (\gamma +1)/2$\\

\hline
\end{tabular}
\label{table:exponents}
\end{table}

An excellent guide to measuring the Hurst parameter can be found in
\cite{hurstguide}. Though the Hurst exponent is well-defined
mathematically, in practice all measurements of it are only
estimations and different techniques, software, or noisy data sets
can produce varying estimates. Even on artificially generated data
with a known Hurst exponent, the different techniques had divergent
measurements of the Hurst exponent and the R/S statistic performed
poorly underestimating $H$ on both generated and real data.

Many may realize that in all of this discussion of self-similarity
and fractals the fractal dimension has not been mentioned once. The
omission is purposeful and due to the convention that almost without
exception, the Hurst exponent is used as the measure of
self-similarity in data traffic research. In any case, the
conversion is not difficult since the fractal dimension $D$ of the
time series is related to the Hurst exponent by

\begin{equation} D = 2 - H
\label{hurstfrac}
\end{equation}

Given equation \ref{hurstfrac} we can see that the typical fractal
dimension of data traffic is around 1.2.

\emph{Nonstationary data methods}

As noted earlier, the previous techniques to measure self-similarity
implicitly assume a stationary signal, however, this is definitely
not the case in Internet traffic where on longer time scales,
non-stationarity due to periodicities such as daily usage patterns
and growth in traffic over time make the traffic data
nonstationary.

A commonly used method of measuring long-range correlations and
self-similarity in nonstationary time series traffic is the use of
detrended fluctuation analysis (DFA), an approach adopted
independently in several papers \cite{DFA1,DFA2,DFA3}. DFA was first
used to measure the long-range correlations in non-coding regions of
DNA and is often used to measure correlations among fluctuations in
physiological or financial time series. In short, DFA is a modified
RMS which calculated the deviation from a trend and long-range
correlation in a time series. To use DFA for a time series $X(t)$ of
length $N$, first calculate the profile $y(t)$ given  by

\begin{equation}
y(t) = \sum^{t}_{i=1} [X(i) - \langle{X}\rangle]
\end{equation}

where

\begin{equation}
\langle{X}\rangle = \frac{1}{N}\sum^N_{i=1}X(i)
\end{equation}

The next step involves separating the signal into $m$ equal sized
non-overlapping segments. In each segment, use least squares
regression to find the local linear trend $\tilde{y}_{t}$ in the
segment and then calculate the detrended profile of the signal,
$y_{m}(t)$ where

\begin{equation}
y_{m}(t) = y(t) - \tilde{y}_{t}
\end{equation}

and finally the detrended rms is calculated as

\begin{equation}
F(m) = \sqrt{\frac{1 }{N}\sum^{N}_{ i=1} y_{m}(t)^2}.
\end{equation}

If the signal has long-range dependence from a $1/f$ spectrum,
$F(m)$ will scale with $m$ as

\begin{equation}
F(m) \sim m^{\alpha}
\end{equation}

$\alpha$ is related to the $1/f$ exponent of the signal by $\gamma =
2\alpha - 1$, which superficially makes it identical to the Hurst
exponent. Give the RMS, the $\alpha$ measured is a second order
measurement of the power law scaling. The Hurst exponent is most
simply extracted by taking the mean value of $\alpha$.

DFA is not the only method for analyzing nonstationary time series
and as some recent studies conclude, is neither always accurate nor
optimum for analyzing nonstationary trended processes. In
\cite{dfacritic} it is mathematically shown that for trended
processes DFA estimates for the Hurst exponent do not converge to an
accurate value and the wavelet method (next section) is recommended
instead to measure the Hurst exponent. Therefore, estimates of $H$
from papers using the DFA method should be viewed with skepticism
and tested against other methods.

\subsection{Wavelet methods}

The final methods usually used to measure self-similarity are
wavelet methods and are often the preferred method for nonstationary
data. In
\cite{multifractal2,multifractal2b,multifractal3,multifractal4,multifractal5},
the logscale diagram method of interpreting Internet traffic is
described. A logscale diagram is created using discrete wavelet
analysis of the signal, where the signal, $X(t)$, is represented as
filtered through a wavelet defined given a timescale $j$ and time
instant $k$ as

\begin{equation}
\psi_{j, k}(t) = 2^{-j/2}\psi(2^{-j}t-k)
\end{equation}

A typical wavelet used in the analyses of these discrete wavelets is
the Haar wavelet. Applying the wavelet transform the signal can be
represented as

\begin{equation}
X(t) = \sum_{k}c_{X}(j_{0},k)\phi_{j_0,k}+\sum^{J}_{j \leq
j_0}\sum_{k}d_X(j,k)\psi_{j,k}(t) \label{waveletedecomp}
\end{equation}

Where $ c_{X}(j_{0},k)$ are called the scaling coefficients, $\phi$
is called the scaling function, and $ d_X(j,k)$ are called the
wavelet coefficients. Wavelet theory will not be covered in great
detail here due to its complexity, however, there are
several useful guides \cite{wavelet1,wavelet2,wavelet3} to the
subject. Each scale increment $j$ represents a scaling of the
timescale of an order $2^j$ and $j$ is commonly termed the octave.
In addition, in \cite{multifractal2b} it is shown for a stationary,
self-similar process that the expectation of the energy $E_j$ that
lies around a given bandwidth $2^{-j}$ around the frequency
$2^{-j}\lambda_0$ where, $\lambda_0$ is the sampling frequency is

\begin{equation}
E[E_j] = E\left[\frac{1}{N_j}\sum_{k}|d_{j,k}|^2\right]
\end{equation}

\begin{figure}[tbp]
\centering
\begin{tabular}{cc}
    \includegraphics[height=2in, width=2.5in]{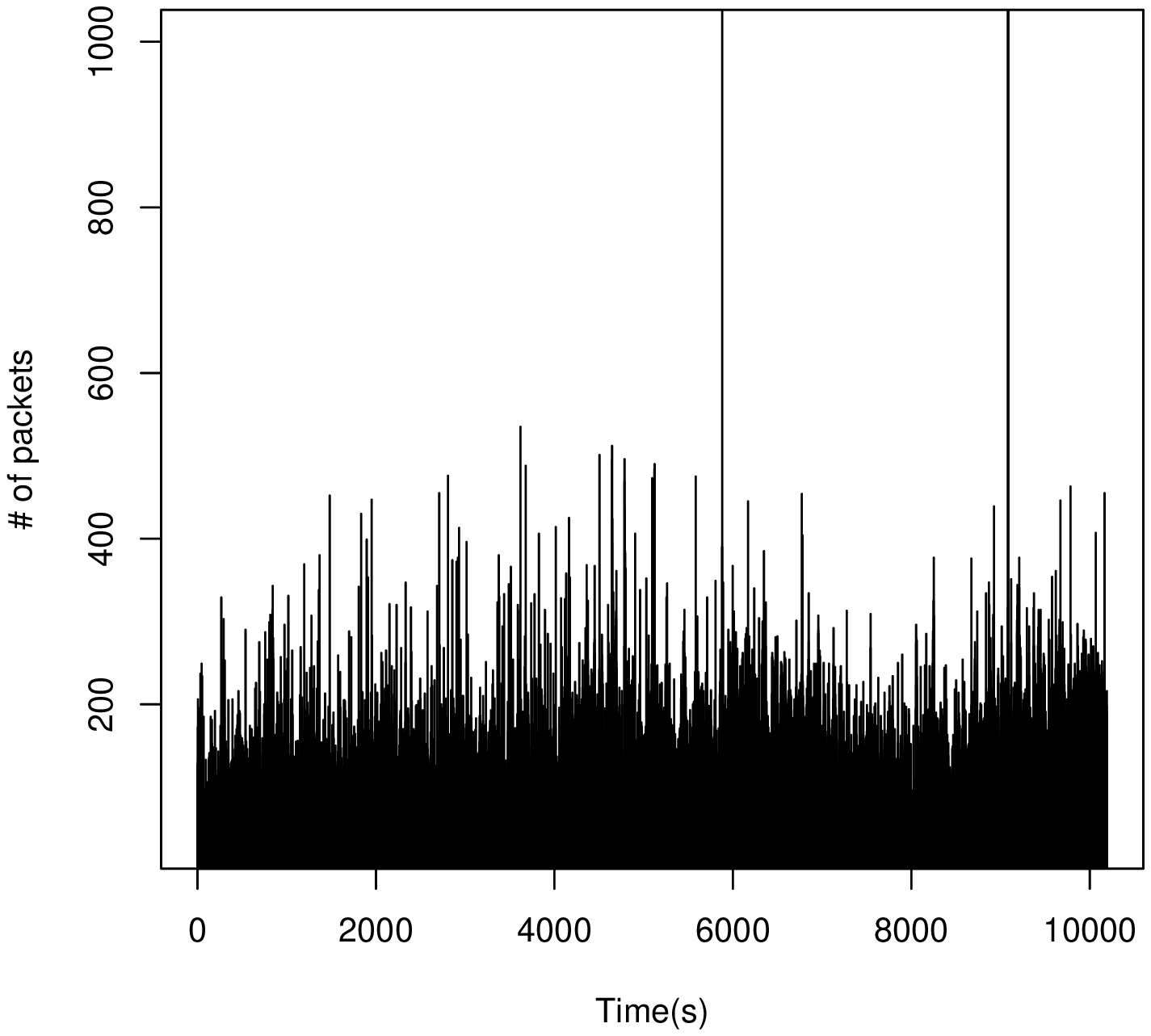}&
    \includegraphics[height=2in, width=2.5in]{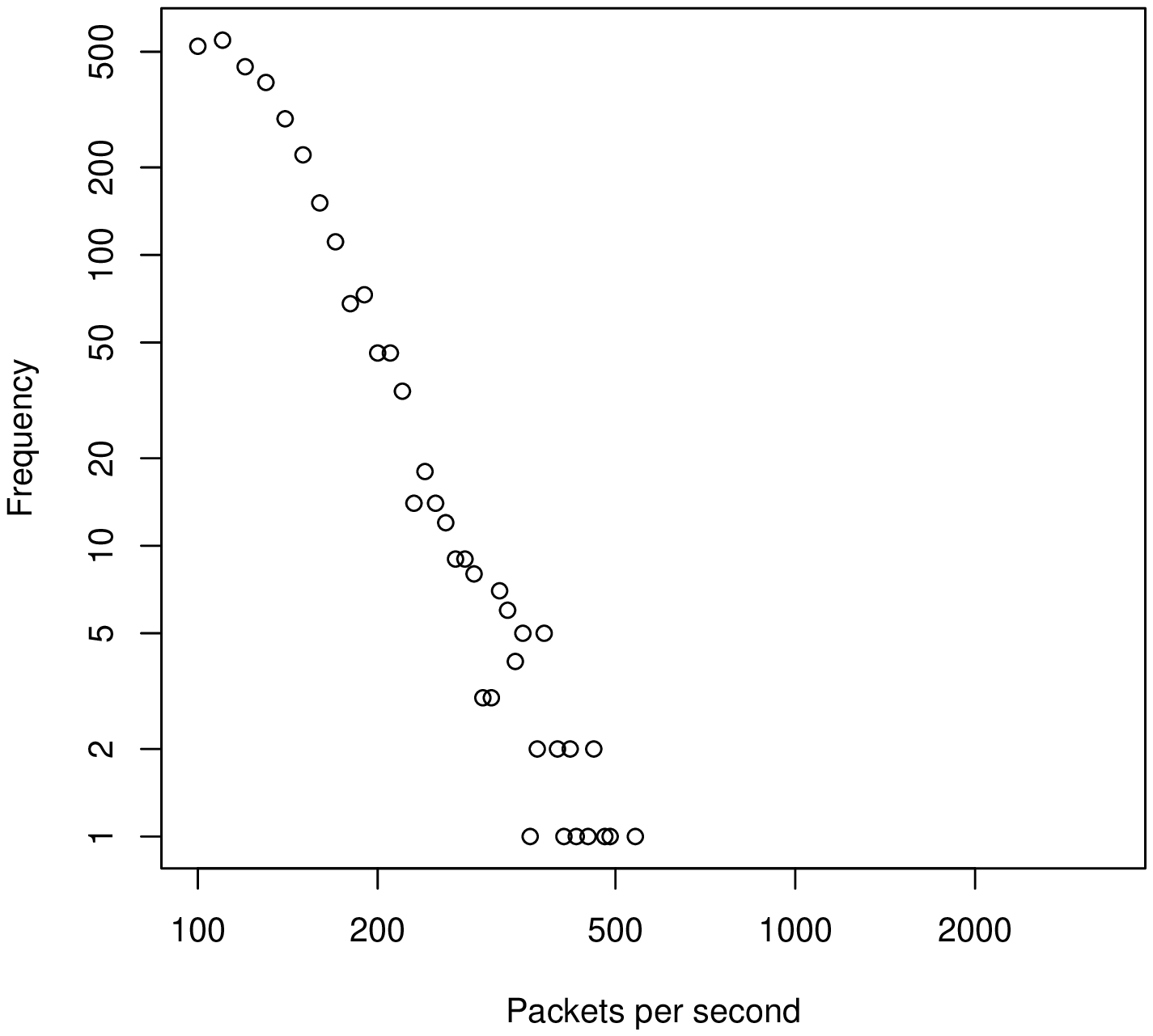}\\
    \includegraphics[height=2in, width=2.5in]{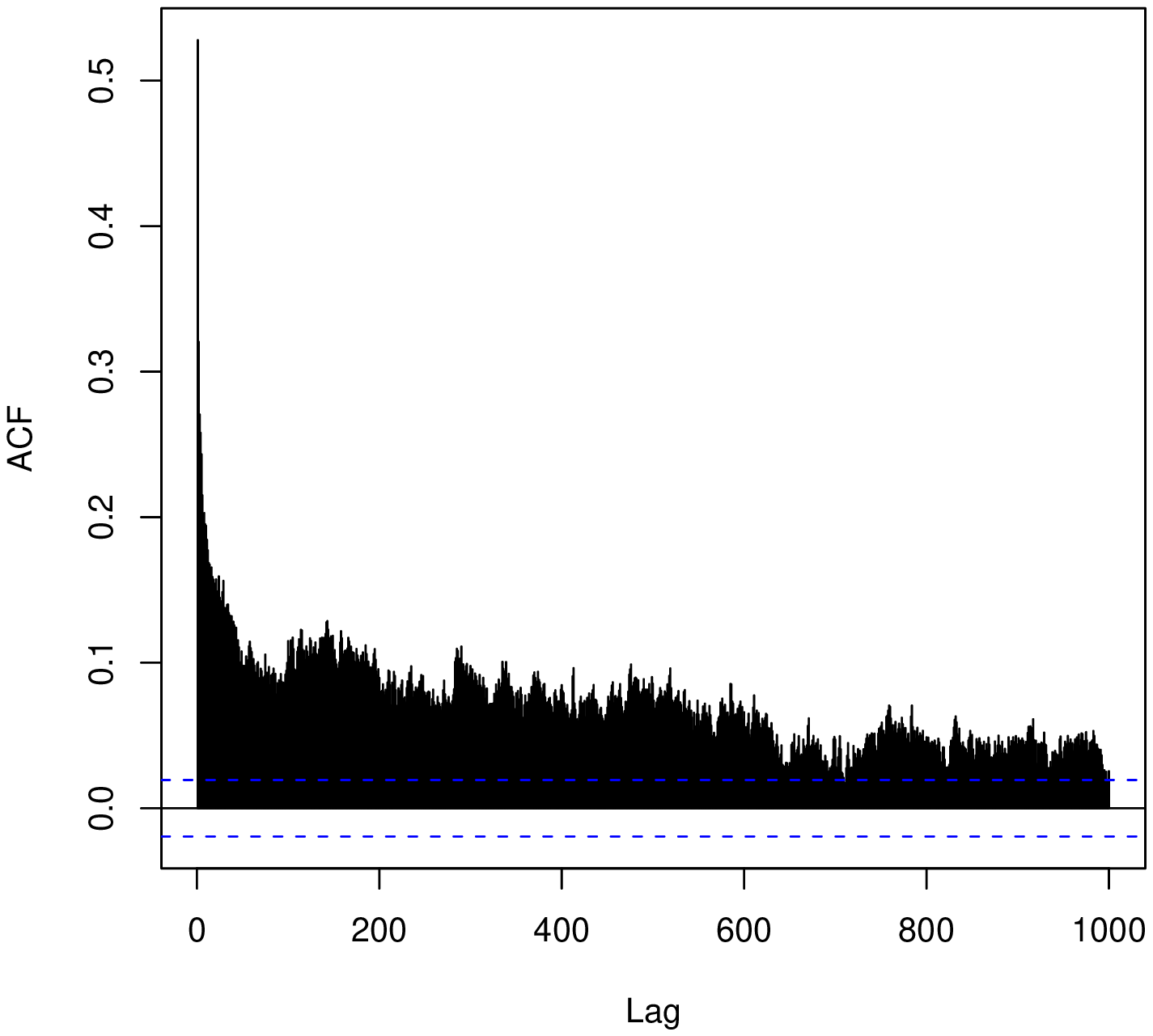}&
    \includegraphics[height=2in, width=2.5in]{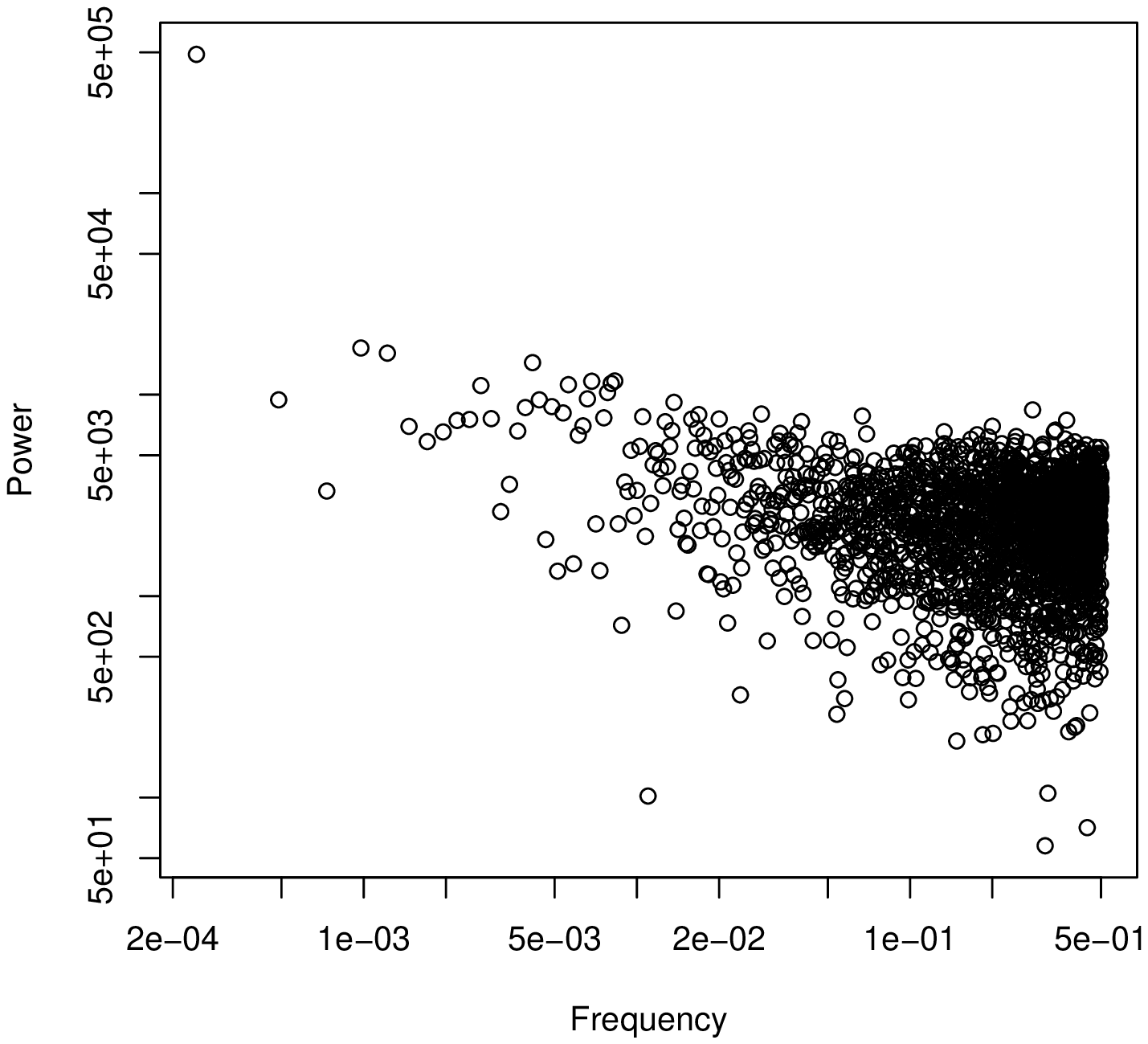}\\
\includegraphics[height=2in, width=2.5in]{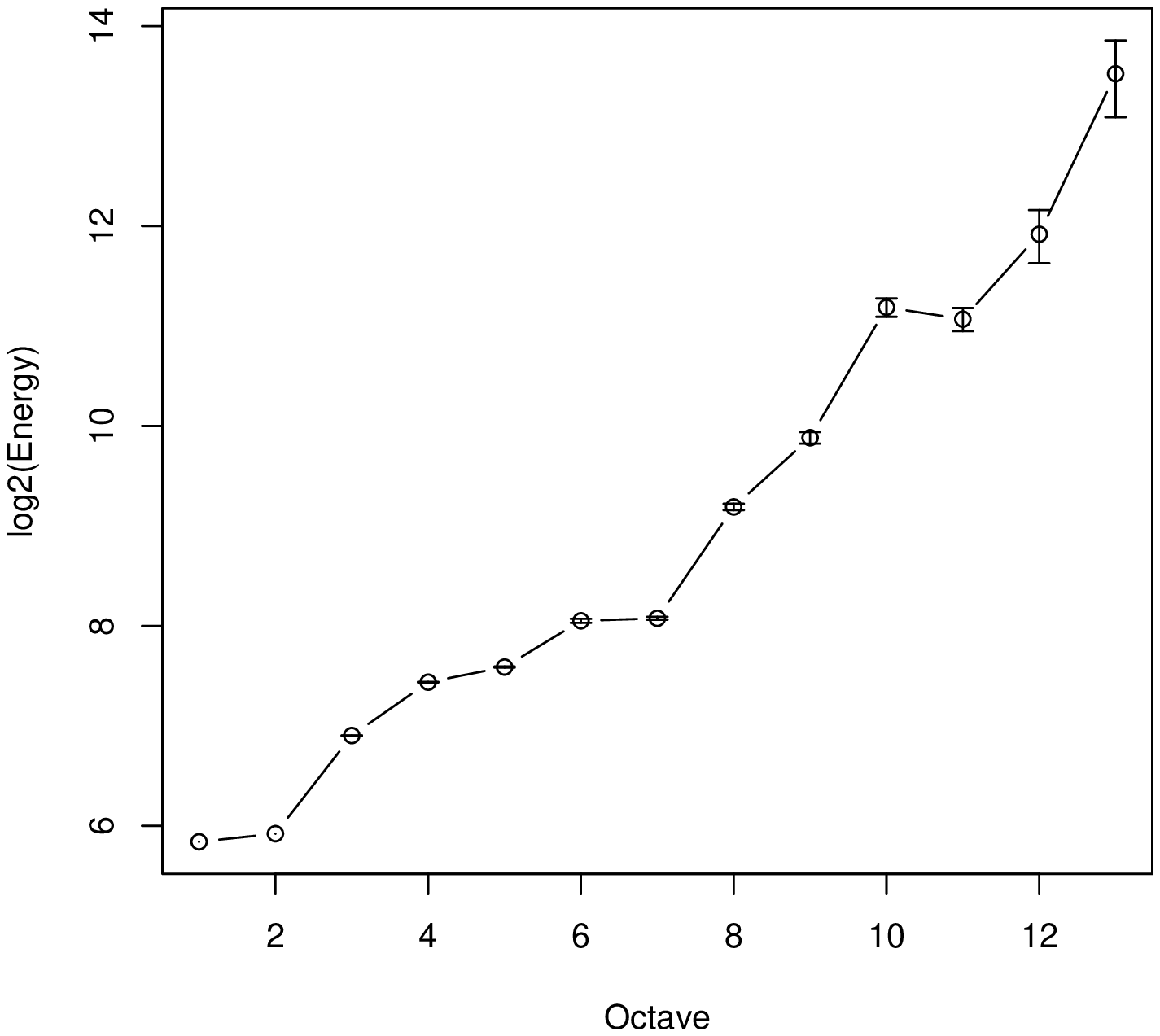}

\end{tabular}

        \caption{A view of a 10,192 second trace of IPv6 6Bone experimental network Internet traffic taken from the WIDE MAWI traffic trace archive Samplepoint-C on July 22, 2005. The data was collected into 1s intervals. The following figures show the packets/s of traffic,
         the distribution of packet arrivals, the autocorrelation of the time series up to a lag of 1000, the $1/f$ noise plot
         of the data trace, and the logscale diagram constructed from wavelet coefficient data based on 100ms bins of packet arrivals and 95\% confidence intervals.
         The Hurst exponent was calculated with the statistical program R with the \emph{fractal} package using the aggregated variance method  estimating H = 0.81.}
\label{traffic}

\end{figure}

Where $d$ are the wavelet coefficients of octave $j$ and $N_j$ is
the number of wavelet coefficients in the octave $j$. Graphing the
log of $E[E_j]$ versus the octave $j$, gives a logscale diagram, an
example of which is the bottom graph in figure \ref{traffic}. In
addition, $E[E_j]$ is also related to the sampling frequency and the
Hurst exponent:

\begin{equation}
E[E_j] = c|2^{-j}\lambda_0|^{1-2H}
\end{equation}

So the logarithm of the expected energy is directly proportional to
the Hurst exponent. In fact, monofractal behavior is indicated by a
linear dependence of $\log{E[E_j]}$ over multiple octaves. The
different scaling regimes can be seen in figure \ref{traffic} by
noting how the curve varies over the octaves 2 to 4 and 8 to 12. There are two important procedures when using logscale diagrams is to always represent the confidence intervals (usually 95\%) for the wavelet coefficient values in each octave. Second, one must make sure to use a minimum variance unbiased estimator instead of least squares regression fitting when estimating $H$ from linear trends between at least three octaves. Appropriate methods for calculating this estimator as well as its variance can be found in \cite{logscalegood1,logscalegood2}.

The second, and often considered more rigorous, method of looking at
changing self-similarity using wavelets is looking at the scaling of
the partition function for each moment of order $q$ over each octave
where the partition function is defined as

\begin{equation}
S(q, j) = \sum_k|2^{-j/2}d_X(j,k)|^q
\end{equation}

The scaling behavior, besides being seen by graphing $log_2 S(q,j)$
vs. $j$ is encapsulated using what is called the structure function

\begin{equation}
\tau(q) = \lim_{j \to -\infty} \frac{\log{S(q,j)}}{j\log{2}}
\end{equation}

if the traffic is exactly self-similar with Hurst exponent, $H$,
then for each $q$, $\tau(q) = Hq - 1$. When more than one scaling
behavior is in the signal, $\tau(q)$ is no longer linear, but
concave and each scaling exponent contributes to its value roughly
according to its relative strength in the signal at the relative
timescale. For details see \cite{multifractal2, multifractal5}

There have been some theories of self-similar behavior in Internet
traffic by traffic engineers. This thesis was first broached and
analyzed in \cite{multifractal0}. In short, it is believed that
self-similarity is generated on long timescales by user and
application needs (see next section on the ON/OFF model) but on
shorter timescales the picture is much less clear. It is possible
that a different self-similarity affected by network or TCP
congestion control considerations is active in this region. Early
papers tried to explain multifractal properties of traffic by a
process known as multiplicative cascades or conservative cascades
\cite{multifractal2, multifractal4, multifractal5b}. The cascade is
mathematically defined as a mass $M$ equally distributed over the
interval $(0,1]$ where the mass is broken up into two new masses,
one with mass $p$ and the other with mass $1-p$, where $p$ is a
fraction of mass defined for the process, and these two new pieces
are broken up again according to the same process ad infinitum. The
multiplicative cascade model was rationalized in the relation to
Internet traffic by describing the encapsulating of flows into
packets and the fragmentation of these packets at the link layer as
a conservative cascade process where the total transmitted data is
conserved but broken down into many different packets. Since this
process occurs over relatively short time scales, it is given as
additional evidence for the cause behind the different scaling at
shorter time scales.

Many questions about the multifractal paradigm, however, were raised
in \cite{multifractal6} which openly criticized some of these claims
and questioned whether multifractal models are necessary and as
proven as they purport to be. In particular, Veitch, Hohn, \&
Abry, while analyzing some of the most common data traces used in
Internet traffic studies, raise questions about the rigor of the
statistical tests used such as logscale diagrams without confidence intervals and the large size of these confidence intervals for some values of the energy at higher octaves. They also raise the point that these tests rely on an
assumption of stationarity in Internet traffic which may not always
be a valid assumption, especially over longer timescales. In the
end, they do not completely rule out multifractals, however, they
raise the point that current statistical tools are not yet fully
developed enough to give a definite answer to existence of
multifractals. Similar comments are made in \cite{bookcritic}
declaring that multifractal patterns may exists but are not to be
seen as an end in themselves and any new model of multifractality
must be matched with a feasible mechanism.

Self-similarity and long-range dependence account for the ``bursty''
behavior of Internet traffic at all time scales. Unlike telephone
traffic, which is Poisson and large spikes are rare deviations from
a mean traffic level and have an exponentially decreasing
probability, burstiness in Internet traffic at almost all-scales has
a non-vanishing probability. This makes traffic management schemes
and infrastructure planning much more difficult from a statistical
standpoint. Sometimes, a scheme known as ``small buffers, high
bandwidth'' \cite{selfsimilar5} is advanced to deal with bursty
traffic to avoid trying to create massive buffers to handle bursts
of traffic. However, there is not yet an easy answer to managing
Internet traffic, especially one with practical use.

\clearpage

\section{Theories on the causes of self-similar traffic}

Once self-similarity was demonstrated in Internet traffic, the next
logical step was to look for the cause. Besides Internet topology models, this is where most of the tension between the different fields has developed and 
caused clashes of theories and methodology. Though some consider the question controversial and unresolved, there are some models which currently have a greater
weight of evidence behind them as checked against real-world
traffic. There are three main theories which will be discussed at
length throughout the rest of the paper.

\subsection{Application Layer Cause: Long-tail ON/OFF sources}
First, there is the most widely known, and only empirically
validated, application layer theory which states that self-similar
traffic is the cause of the behavior or users. This was first
elaborated in \cite{selfsimilarcause1, selfsimilarcause2}. This
theory models the traffic on the Internet as large number of ON/OFF
sources with identical duration distributions, an idea earlier
broached by Mandelbrot \cite{selfsimilarcause2b}. The ON/OFF
sources, which reflect flows, are superimposed traffic sources that
alternate in ON and OFF periods according to a power-law
distribution. Though this is not usually explicitly mentioned, this
model is extremely close to modeling a large number of flows with
long-tails dominated by a few elephant flows found in actual
traffic measurements. In \cite{selfsimilarcause1,selfsimilarcause2}
they give evidence both from theory and observation that many
superimposed ON/OFF sources behave to a limit as fractional Brownian
motion and can account for the self-similarity seen in overall
Internet traffic. Specifically, the Hurst exponent of the traffic of
the superimposed is calculated by
\begin{equation}
H = \frac{3-\mbox{min}(\alpha_{ON}, \alpha_{OFF}))}{2}
\label{onoff}
\end{equation}

where $\alpha_{ON}$ and $\alpha_{OFF}$ are the power law exponents
of the ON and OFF distribution times respectively. ON/OFF flows with
long-tail duration are not merely a theoretical abstraction but have
a basis in the access of files over the Internet. Internet file
sizes have been shown by several studies \cite{selfsimilarcause4,
selfsimilarcause5,selfsimilarcause5b,selfsimilarcause5c} to be at
least long-tailed though there is dispute over whether the
distribution more closely fits a power law, lognormal, or double
Pareto.

Crovella \& Bestavros and Crovella, Park \& Kim
\cite{selfsimilarcause3, selfsimilarcause3b, selfsimilarcause3c}
extend the model to explain the self-similar nature of TCP and web
traffic. They base their studies on long-tailed file sizes with a
power-law exponent of about 1.2 (implying a Hurst exponent of about
0.9 from equation \ref{onoff}). They also found that in general,
reducing the tail of the file length distribution by increasing the
power-law exponent, lowers the Hurst exponent as expected in
\cite{selfsimilarcause1,selfsimilarcause2}. This theory of ON/OFF
sources, and its variants, has become the most dominant explanation
for self-similar network traffic in most network engineering papers and has the largest theoretical and empirical backing.
It also is a frequently used model for simulating data traffic in
Internet simulations.

The second theory, discussed more in detail in the next section,
considers origins of the self-similar traffic at the transport
layer. In particular, it looks at possible effects the TCP
congestion control algorithms may have on network traffic given the
feedback and collective behavior it can engender among multiple
traffic sources over the same path. The third main theory, discussed
in the section on phase transition models of traffic congestion is
that which connects self-similarity to critical phenomena in data
traffic near the transition point from free flow to congested
traffic. This is the model currently most favored amongst physicists
and underlines many of the models of data traffic that will be
described later. In addition to describing these models, the
insights and flaws of each will be described in detail.

\subsection{Transport Layer Cause: TCP Congestion Control}

As discussed earlier, there has been interest in whether
self-similar traffic can find its causes in the congestion control
of TCP rather than at the application level. This has not been a
generally accepted or verified cause or contributor to
self-similarity, at least on long time-scales. Veres and Boda
\cite{TCPchaos1} first brought up the conjecture that since the
assumption TCP stochasticity or predictable periodicity itself in
these equations is highly flawed, TCP throughput then cannot be
reduced to a closed form equation, and that TCP instead exhibits
deterministic and chaotic behavior. In addition, most analyses look
at TCP only at the single link level instead of treating it as a
network dependent entity given congestion control. They based their
papers on simulations that showed self-similarity, sensitivity to
initial conditions, strange attractors, and stable periodic orbits
appeared. Fekete and Vatay \cite{TCPchaos2} also used simulations to
show that the interaction of TCP with buffers in routers can also
cause chaotic behavior in the TCP flows. They simulated the
interaction of $N$ different TCP flows with a buffer that had the
capacity to hold a fixed number of packets. They show that the
backoff algorithm of TCP caused by lost packets can cause power-law
behavior in packet interarrival times and chaotic dynamics given the
(buffer length)/($\#$ of TCP packets) ratio is below a critical
value of 3.

Similarly, Hag\`{a} et. al. \cite{TCPchaos4} used simulations to
recreate self-similar traffic and long-range dependence by the
interactions of multiple TCP flows at a buffer on a central router
connecting three different hosts. They assume an effective loss
packet loss rate (real packet losses and RTT exceeding the allowable
timeout period) but have an infinite sized router buffer so there is
no real packet loss but a large effective packet loss due to
timeouts. This model can produce self-similar traffic with $H$ =
0.89 without any assumptions of ON/OFF distributions or file sizes
beyond a constant TCP flow size of 1000 packets. The only extra
assumption is stochastic source and destination of TCP flows among
the three hosts.

These methods, particularly relying on simulations or analytical
reasoning and not actual Internet traffic trace methods, have been
heavily criticized, especially by those favoring an application
layer explanation for self-similarity. In \cite{TCPchaos8,
TCPchaos9}, a more comprehensive TCP model is developed accounting
for both the backoff phase and congestion avoidance phase after slow
start. Their argument is that TCP can possibly generate correlation structure
and possibly self-similarity but only short timescales (up to
1024xRTT for high packet loss) and not arbitrarily long time scales
which characterize most self-similar traffic. Their models shows
that at low packet loss rates the correlation structure is dominated
by congestion avoidance after slow start while the exponential
backoff governs the correlation structure at high loss rates. In
\cite{TCPchaos5}, the authors focus on short TCP flows and thus only
model slow start and backoff. They again show self-similarity at
large enough packet loss rates but though the long-range dependence
that is also present would connote infinite variance, the TCP based
self-similarity only extends over certain short timescales. Thus the
authors dub this self-similarity pseudo self-similarity since its
timescale is relatively limited. Veres et. al. answer these
criticisms in \cite{TCPchaos10}. Here they concede that though TCP's
congestion control may not by itself be the cause of LRD in Internet
traffic, they show through data, simulation, and mathematical
arguments that TCP's congestion control suite may propagate
self-similar traffic along its path if it encounters a bottleneck
that limits its send rate and has self-similar traffic. Therefore
even if TCP can't create the full self-similar effect, it may be
responsible for propagating the self-similarity far beyond the
traffic it originated at.

In a variation of the above research, Sikdar and Vastola
\cite{TCPchaos6, TCPchaos7} give a model where self-similarity and
long-range dependence emerge from the dynamics of a single TCP flow
instead of multiple flows. They model a single TCP flow as the
superposition of $W_{max}$ ON/OFF processes where $W_{max}$ is the
maximum window size advertised by the receiver. This is similar to
the earlier ON/OFF model but they  show that for higher packet loss
rates, a higher Hurst exponent and more self-similar traffic is
generated. In addition, there are other papers detailing possible
mechanisms by which TCP congestion control can give rise to
self-similarity in Internet traffic.

In the end, TCP congestion control has not emerged as a favored
cause of self-similar traffic. Perhaps it is a strong or even
dominant factor of self-similarity at relatively short timescales
but is likely not the cause of the pervasive self-similar traffic
described in most papers. One final note not connected to the idea
of self-similarity is that TCP is one of the few ways that traffic
actively couples to the network topology via the dependence of
throughput on RTT. Though most latency in networks is likely caused
by congestion and other network conditions, given similar bandwidths
and delays, the average shortest path (number of hops) in topology
can affect average RTT as shown in \cite{hops}. In summary, TCP
being the dominant protocol on the Internet is one the main
determinants of the traffic dynamics. However, TCP is a complicated
and feedback driven protocol whose actions can only be partially
estimated using analytical or stochastic models. The TCP protocol
will definitely hold promise in the future for those looking for
more intricate complex phenomena or pattern formation in Internet
traffic dynamics.

\section{Network Layer Cause: Theories of Phase Transitions and Critical Phenomena in Networks}

The most active area of work by physicists in the research of
network dynamics is a group of research which merges the new
insights of Internet traffic behavior with the mature and
well-tested tools of statistical mechanics and critical phenomena.
Similar to papers written on vehicle traffic
\cite{vehicle1,vehicle2,vehicle3} these papers have analyzed the
onset of congestion in networks as a phase transition from a
free-flow to congested state determined by a critical parameter. In
fact, an explicit comparison was given in \cite{vehicle4}. The
papers, in general, deal with three broad, though sometimes
overlapping, themes regarding the onset of congestion. First, are
the papers that analyze the onset of congestion as a function of the
packet creation rate for various topologies and also whether the
self-similar structure of traffic can be reproduced in these models.
Second, are models primarily concerned with investigating the rise
of self-organized, emergent phenomena in networks in the critical
state and linking the studies of congestion with the study of
self-organization in general. Finally, are many papers who
investigate how different routing strategies can delay or affect the
onset of congestion. The papers of the last category often overlap
with the first. Below I will describe some of the most often quoted
papers. A more comprehensive list and reference of papers is given
in table \ref{routingtable} for those wanting to delve into the
topic in more detail. In the papers described, the critical
parameter is typically the packet creation rate. This has different
symbols depending on the paper but here we will describe it as
$\lambda$. Finally, in the next section I will note many common, and
unfortunately many times accurate, criticisms and problems with
these models.

Papers by physicists investigating congestion first concentrated on
the onset of congestion as a critical phenomenon and possible links
between this and the self-similar nature of Internet traffic. With
few exceptions, these papers focus on the link or network layer
dynamics (IP) as the source of critical phenomena in Internet
traffic. One of the first papers to deal with a phase transition
model of Internet traffic was by Csabai in 1994 \cite{netcongest1}.
In this paper, Csabai noted the presence of a $1/f$ power spectrum
for the RTT times for pings between two computers where the fitted
slope is -1.15 (about an $H$ of 1.08). He also is among the first to
compare Internet data traffic with vehicle traffic \cite{vehicle4}.
It must be noted that the RTT from ICMP echoes is not always
equivalent to the RTT in TCP since many gateways give preferential
forwarding to TCP packets. Also, this power law spectrum based off
of ICMP echoes is different from the overall traffic whose
self-similarity was discussed earlier.

Takayasu, Takayasu, and Sato \cite{netcongest2} followed up with a
similar study where they also note the $1/f$ distribution of RTT for
ICMP pings between two computers if there are many gateways on the
route between them, likely because of consecutive jamming due to
filled buffers. For a short route, their echo replies are
distributed $1/f^{2}$ at low frequencies and as white noise at
higher frequencies ($f > 10^{-4}$). They extend the analysis though
to include a theoretical derivation of the behavior of network
traffic taking into account a simple topology. They disregard loops
and use the theoretical topology of a Cayley tree where gateways are
sites and cables are links. A contact process (CP) is modeled where
empty sites are considered jammed gateways and filled sites
(particles) are considered un-jammed gateways. A jammed gateway has
a probability $p$ of becoming un-jammed if it is adjacent to an
un-jammed gateway (particle reproduction) and an un-jammed will
become jammed with an independent probability $q$ (particle
annihilation). An un-jammed gateway will do neither with probability
$1 - p - q$. In analyzing the simulation, they assume that the
number of un-jammed gateways over time is equivalent to the
distribution of RTT. They derive a power-law result from the CP
process which shows the distribution of jammed sites over time
follows a $t^{-\alpha}$ power law distribution with time when a
parameter $\delta = 1 - p/q$ equals 0 and that this power law yields
$1/f$ noise for the conditions $0 \leq \alpha \leq 1$. A comparison
of ICMP echo RTTs to earthquake aftershocks is made by Abe and
Suzuki \cite{netcongest2x} who fit the RTT from pings in Internet
traffic to a statistical distribution that is similar to Omori's Law
which models the arrival of aftershocks from an earthquake.

A similar hierarchical tree topology is used to investigate critical
behavior for data flow by Arenas, Di\`{a}z-Guilera, and Guimer\'{a}
\cite{netcongest2c}. They derive a mean-field theory solution for
the critical packet creation density and also show that most
congestion occurs at the root of the tree and the first level of
branching. Power-law scaling of the total number of the packets in
the system is observed near the critical point $\lambda_{c}$.

Takayasu \& Takayasu later expand on a theory of self-similarity in
Internet traffic as a critical phenomenon in \cite{netcongest2a,
netcongest3, netcongest2b} In \cite{netcongest3} Takayasu, Takayasu,
and Fukuda describe what they believe is a phase transition in the
flow of overall Internet traffic. They separate data traffic into
500 s bins and take autocorrelations of each bin comparing the
correlation length in seconds with the mean traffic density. The
correlation length increases with traffic flow density until a
critical density $\lambda_{c}$ = 500 kbytes/sec where the
correlation length begins decreasing again. They associate this with
a second order phase transition in the flow where there is a
transition from free to congested flow. When they consider any flow
above 300 kbytes/sec as ``congested'' around the critical point they
can show power law scaling of lengths of congestion times confirming
the critical nature of the phenomenon. In \cite{netcongest2b}, the
same authors theorize that the critical nature of traffic measured
in Ethernet networks is due to the Ethernet collision detection
management algorithm (CSMA/CD) which employs an exponential backoff
algorithm on detection of an Ethernet frame collision that is
qualitatively similar to the congestion backoff mechanism described
in TCP. They show that a binary backoff algorithm can generate $1/f$
traffic distributions at the critical point.

Most of the other prominent papers in the first category follow in
the tradition of the first Takayasu paper describing phase
transitions using  modeled networks in simulations to infer a shift
in dynamics above a certain packet creation/flow threshold
\cite{netcongest4b,netcongest4c,netcongest5b,netcongest4,
netcongest5,netcongest6b,netcongest6b2,netcongest6c, netroute6}. The
threshold can be changed either by topologies, which are more
efficient with small world networks such as random graphs or
scale-free graphs than lattices or Cayley trees, by changing the
proportion of nodes that can generate (versus only route) traffic,
or by routing strategies such as in \cite{netcongest4,netcongest6c}.
In \cite{netcongest5, netcongest6b, netcongest6b2}, mean field
methods are used to calculate the onset of congestion for a 2D
lattice at

\begin{equation}
\lambda_{c} = \frac{2}{pL}
\end{equation}

where $p$ is the proportion of nodes that can generate packets and
$L$ is the length in nodes of a side of the lattice. All of these
papers purport to show $1/f$ distributions of packet travel times
which they link to self-similar traffic. A paper looking at the
problem from a different angle by Moreno, Pastor-Satorras,
V\'{a}quez, and Vespignani \cite{netcongest6d} approached the
problem by looking at what average traffic density in the overall
network could lead to a spread of congestion across all nodes and
the collapse of the network. This is a related viewpoint on the
cascading router failures and percolation models that have been
studied on scale-free topologies \cite{cascade1,cascade2,cascade3}
which links cascading failures not just to topological sensitivity
of certain hubs but also the traffic levels in the network.

In the second category, are papers largely concerned not with the
value of the critical parameter but with emergent phenomena
themselves. One of the earliest papers hinting at this was a study
by Barth\'{e}lemy, Gondran, and Guichard \cite{netcongest8b}.
Borrowing techniques from nuclear physics, they studied the
eigenvalue distribution and eigenvectors of the traffic correlation
matrix of 26 routers and 650 connections in the Renater computer
network for two weeks of traffic data. Their technique used random
matrix theory to compare the eigenvalue distribution of the
correlation matrix of Renater traffic fluctuations to that of a
control random matrix. The traffic fluctuations in an interval
$\tau$ in the traffic between source $i$ and destination $j$ were
defined as

\begin{equation}
g_{ij}(t) = \log\left[\frac{F_{ij}(t+\tau)}{ F_{ij}(t)}\right]
\end{equation}

and the correlation between connections $ij$ and $kl$ is defined as

\begin{equation}
C_{(ij)(kl)} =\frac{\langle g_{ij}g_{kl}\rangle- \langle
g_{ij}\rangle\langle g_{kl}\rangle }{\sigma_{ij}\sigma_{kl}}
\end{equation}

They found that the largest eigenvalues were much larger than the
largest eigenvalues of a similar rank random traffic matrix whose
flows have a mean of 0 and unit variance. Also, the largest values
of the eigenvector for the largest eigenvalue correspond to the most
highly correlated routers. These results all indicated
spatiotemporal correlations among the routers in the network that
deviated from traffic defined purely by a stochastic process.

Among the most consistent researchers to address the emergent
phenomena question directly are Yuan and Mills \cite{netcongest7,
netcongest8, netcongest9, netcongest10} who make a persuasive case
that emergent phenomena in networks could go beyond the simple onset
of congestion in simple network topologies and only treating packets
at the network (IP) layer. The main themes of their papers are
measuring spatiotemporal patterns that emerge in larger networks.
The main features they add lacking in many other models are size
(more nodes), more realistic topologies, as in \cite{netcongest9},
where their network includes four levels of hierarchy in tree
structure, and modeling of transport (TCP) level effects such as
congestion control \cite{netcongest9}.  In their first paper
\cite{netcongest7}, they use a simplified topology of a 2D cellular
automata (CA) with all nodes as hosts and routers. The state of a
router on the CA is defined by the number of packets in its queue
and it ``transitions'' by passing off packets given the state of the
queues of the surrounding cells. The traffic sources can originate
in any node and are modeled as ON/OFF sources as in
\cite{selfsimilarcause1, selfsimilarcause2}. Packets are routed via
a full routing table. In addition, they model the systems with three
types of congestion control algorithms: no congestion control, a
congestion control that stops transmitting above a threshold RTT per
hop to the destination, and a TCP-imitating congestion control that
includes slow start and congestion avoidance.  Their main results
use the TCP-imitating congestion and produce power spectrums of the
time series of the number of received packets at a given node for
various sample time lengths and network sizes. In general, they find
that increasingly longer sample times diminish the correlations and
long-range dependence measured in the power spectrums but increasing
the network size increases the correlations over both time and
space. Comparing smaller networks to similar sized subgraphs in
larger networks shows that the subgraphs exhibit stronger
correlations and they deduce that large network sizes can allow for
wider coupling and self-organization. They also surmise larger
networks may be more predictable because congestion is stable over
longer time scales.

In subsequent papers, this idea is developed further. In
\cite{netcongest8}, the authors do an analysis by creating a weight
vector for each node that is constructed from the components of
eigenvectors derived from the correlation matrix as in
\cite{netcongest8b}. Yuan and Mills create a technique to analyze
simulated networks of a larger size. They define flow vectors,
$x_{i}$,  where $i$ ranges from 1 to N where N is the number of
nodes with N components each component $x_{ij}$ representing the
flow from node $i$ to node $j$ during a sample interval. They then
create a normalized flow vector by normalizing each element of each
vector for the entire sample time including all intervals where the
normalized vector is

\begin{equation}
f_{ij} = \frac{x_{ij} - \langle{x_{ij}}\rangle}{\sigma_{ij}}
\end{equation}

They analyze the eigenvector of the largest eigenvalue of the
correlation matrix among all normalized flow pairs over time and use
the elements of the subvectors of this eigenvector to create N $S$
vectors where $S_{ij}$ is the relative contribution of node $i$ to
node $j$ in terms of traffic correlations. Performing simulated
traffic on a 2D topology with ON/OFF sources with Pareto distributed
ON times and TCP congestion control they observe complex fluctuation
of the largest eigenvalue over times as well as correlated traffic
between certain nodes over time though they note their largest
eigenvalues tend to be smaller than those in \cite{netcongest8b}.
They also raise the point that during congested critical states,
taking a sample of a few nodes (or routers) may give a better and
more overall cohesive picture of the entire network if sampling all
nodes is infeasible. This is mainly due to the increased
spatiotemporal correlations in congestion. Also, longer sampling
time windows tend to reduce the visibility of correlations in
traffic.

In \cite{netcongest9}, they continue the same research based on the
eigenvalue method but using a four-tier (backbone router, subnet
router, leaf router, and source hosts) hierarchical network to model
the actual AS-level and below topology of the Internet. They also
use the measures of spatiotemporal correlation to find both hotspots
and show that distributed denial of service (DDoS) attacks can cause
large-scale effects beyond the target router by disturbing traffic
flows in other correlated routers in the network. They suggest
methods of analyzing network-wide phenomena using small samples of
nodes and possibly detecting DDoS attacks by the signatures of
large-scale perturbations in correlated network traffic. In
\cite{netcongest10}, they return to the 2D CA formalism but
investigate spatiotemporal dynamics using wavelets and logscale
diagrams over varying average packet creation rates, congestion
control protocols, and average flow durations. They looked for
causes of LRD at the application level (file size distribution),
transport level (congestion control type) and network level (varying
the rate of ON/OFF sources and network size). They found that LRD
emerged on wide times scales with long-tail distributions of file
sizes, an increasingly large network size, or Pareto distributed
ON/OFF source times but only emerged on limited time scales when
only the type of congestion control was varied. Though they
acknowledge the limits of their model they suggest that most LRD
emerges due to interactions in the network layer or possibly
file-sizes in the application layer. Like \cite{TCPchaos8,
TCPchaos9, TCPchaos5} they suggest TCP congestion control plays only a
limited part in the emergence of LRD. Yuan et. al.
\cite{netcongest11} closely replicate the results of
\cite{netcongest8} except they compare visualization of the largest
eigenvalue over time with the information entropy of the weight
vectors. They find the eigenvalue more clearly shows the change in
correlation structure over time. There are also some very
interesting spatiotemporal plots of router congestion over time in
\cite{netcongest12} showing pattern formation in the temporal
congestion among routers in a 1D cellular automaton model.

Other researchers, still following the theory that the onset of congestion could be considered a critical phenomenon,  began investigations on possible
new routing strategies that could help extend the tolerance of a
network to congestion. In short, all of the proposed routing
strategies aim to be an improvement over current state Internet
routing where routers use a global router table and shortest path
metrics to route packets. In particular, these papers show that the
geodesic on the networks between two points defined solely according
to a graph shortest path are not always the best routing paths in
real traffic conditions. The newly proposed routing strategies tend
to explicitly take into account traffic and/or queue conditions at
neighboring routers or use different topological measures such as
betweenness \cite{netroute2, netroute1} in order to redefine the
shortest path metrics and packet routing strategies. Besides random
walk routing, the most common routing heuristics are next nearest
neighbor (NNN) \cite{netroute4,netroute4b,
netroute4b2,netroute4c,netroute4d} based on delivery to the
destination if it is an adjacent node or  either random walking or
bias to higher degree nodes otherwise (except a packet cannot travel
to a node it has just left). Typically it is shown that NNN is
superior to random routing. Preferential next nearest neighbor
(PNNN) \cite{netroute5, netroute5also, netroute5also2,
netroute5also3} slightly alters NNN by explicitly taking into account
node degree in routing according to a parameter $\alpha$ which
creates a preference distribution for nodes of a degree $k$
according to the relation $k^{\alpha}$, or a similar relation. In
\cite{netroute5a} traffic congestion at nodes is also taken
explicitly into account. Typically, there is a purported value of
$\alpha$ which minimizes average packet travel time and tolerates a
higher packet creation rate for the onset of congestion.

Many of these papers, with some exceptions such as
\cite{netcongest5}, do not show how these new routing methodologies
should compare against the current shortest-path full routing table
the Internet uses for routing. What would the value of a random walk
or walk based on the degree of an adjacent router add to network
routing infrastructure and performance? Or is it worse than the
current system (which most comparisons suggest)? The basic idea of
the geodesic between source and destination depending on traffic
conditions is a very interesting proposal but how could it be
implemented in practice? These questions should provide fertile
ground for future research and cross-disciplinary collaboration.

There have been some papers proposing feedback based routing in the
network engineering literature (for example \cite{netroute7}),
however, they are not related to similar research in physics and
this line of research by physicists is often not looked upon highly
by the network engineering community as will be discussed in the
next section. One final note from equation \ref{basiceq} is that
since all of these models use the packet as the basic unit, the
concept of the relationship between data throughput and packet size
shows that apart from topology changes or new routing,  as stated
earlier, one easy way to reduce the packet creation rate on a
network is to increase the average packet size. Since throughput is
an important measure in the function of the Internet, future
measurements and experiments on packet creation and congestion
should thoroughly account for this.
\clearpage

\scriptsize{
\begin{longtable}{|p{0.5 in}| p{0.75 in}| p{0.75 in}| p{0.75 in}| p{1.5 in}|}

\hline
Citation&Topology&Host/Router Distribution&Packet Creation Distribution&Routing Strategy\\
\hline
\cite{netcongest2c}&Cayley Tree&All nodes are both&uniform distribution; fixed probability&full routing table - shortest path\\
\cite{netcongest2a}&Cayley Tree&Hosts on perimeter&uniform distribution; fixed probability&full routing table - shortest path\\
\cite{netcongest4}&2D square lattice&Hosts on perimeter&Poisson process of rate $\lambda$&1) deterministic method where packets are routed to nodes who have  received least packets, 2) probabilistic method where particles choose node biased against nodes having handled more packets\\
\cite{netcongest4b}&2D square lattice&All nodes are both&uniform distribution; fixed probability&proximity of neighbors to destination, inverse ``temperature'' thermal agitation, and repulsion with sites with filled buffers\\
\cite{netcongest4c}&2D square lattice&Top row nodes are hosts; bottom row destinations&uniform distribution; fixed probability&random walk\\
\cite{netcongest5b}&2D square lattice&All nodes are both&uniform distribution; fixed probability&full routing table - shortest path\\

\cite{netcongest5}&2D square lattice + extra links&All nodes are both&Poisson process of rate $\lambda$&using a routing table and lowest queue consideration in neighbors; use two types of routing tables: a full routing table with all paths and a partial table with paths only up to a distance m from each node\\
\cite{netcongest6}&2D square lattice&random nodes of probability $p$ are hosts&uniform distribution; fixed probability&full routing table - shortest path \& congested node avoidance\\
\cite{netcongest6b}&2D square lattice&random nodes of probability $p$ are hosts&uniform distribution; fixed probability&full routing table - shortest path; $\lambda$ moderated by congestion \& congested node avoidance\\
\cite{netcongest6b2}&2D square lattice&All nodes are both&Poisson process and long range dependent distribution&full routing table - shortest path \& congested node avoidance\\

\cite{netcongest6c}&random, regular, Cayley tree, scale-free&All nodes are both&uniform distribution; fixed probability&full routing table - shortest path, and biased against high degree or betweeness nodes\\
\cite{netroute6}&1D chain, 2D lattice, Cayley tree&All nodes are both&uniform distribution; fixed probability&full routing table - shortest path\\

\cite{netcongest6d}&scale-free&All nodes are both&initial load created on edges by uniform distribution&N/A\\
\cite{netcongest7}&2D cellular automata&All nodes are both&Poisson duration ON/OFF sources&full routing table - shortest path\\
\cite{netcongest8}&2D square lattice&two tiers: one hosts, one routers&Pareto duration ON/OFF sources&full routing table - shortest path\\

\cite{netcongest9}&four tier hierarchical network&tier four (lowest level) sources and receivers&Pareto duration ON/OFF sources&full routing table - shortest path; different forwarding capacities on each tier\\
\cite{netcongest10}&2D cellular automata&two tiers: one hosts, one routers&both Poisson and Pareto duration ON/OFF sources&full routing table - shortest path; different forwarding capacities on each tier\\
\cite{netcongest11}&2D square lattice&two tiers: one hosts, one routers&Pareto duration ON/OFF sources&full routing table - shortest path; different forwarding capacities on each tier\\
\cite{netcongest12}&1D chain &Fixed number of hosts at random positions on chain; inter-host spacing is buffer length&uniform distribution; fixed probability&right to left diffusion based off of max routing speeds and buffer sizes\\
\cite{netroute1}&scale-free&All nodes are both&uniform distribution; fixed probability; one time packet creation t=0&global routing table; shortest path and congestion in neighbor nodes\\
\cite{netroute2}&scale-free based on Internet AS map&All nodes are both&uniform distribution; fixed probability&global routing table\\
\cite{netroute3}&2D lattice, scale-free&All nodes are both&fixed number of packets per time step&global routing table; shortest path biased against high betweeness nodes\\
\cite{netroute4b, netroute4b2}&scale-free, web like&All nodes are both&uniform distribution; fixed probability&NNN\\
\cite{netroute4c, netroute4d}&scale-free, web like, random grown tree&All nodes are both&fixed number of packets per time step&NNN, random walk, 1-2 distance awareness\\
\cite{netroute5}&scale-free&All nodes are both&fixed number of packets per time step&PNNN\\
\cite{netroute5also}&scale-free&All nodes are both&fixed number of packets per time step&1) routes to neighbors biased towards higher degree nodes; 2) also adjusted for congestion in neighbors\\
\cite{netroute5also2}&scale-free&All nodes are both&fixed number of packets per time step&PNNN\\
\cite{netroute5also3}&scale-free&All nodes are both&fixed number of packets per time step&PNNN\\
\cite{netroute5a}&random, static scale-free, BA scale-free&All nodes are both&uniform distribution; fixed probability&full routing table - shortest path \& congested node avoidance\\
\cite{netroute5b,netroute5c}&2D square lattice + extra links&All nodes are both&Poisson process of rate $\lambda$&full routing table - shortest path \& congested node avoidance\\
\hline
\caption{A general view of the statistical mechanics
congestion and routing models discussed in the paper.}
\label{routingtable}

\end{longtable}
}
\normalsize
\section{Criticisms of Various Approaches to Self-Similarity}

Though the physics literature on congestion and critical phenomena
in networks is becoming increasingly sophisticated and adept at
``reproducing'' self-similar patterns seen in Internet traffic,
there have been several valid criticisms, particularly from the
network engineering community, that the methodologies may reproduce
observations but do not take into account the actual workings of the
Internet in detail \cite{netcritic1, selfsimilar7, netcritic3, DFA1,
netcriticnew3,netcriticnew4,netcriticnew5,netcriticnew6,netcriticnew7,netcriticAMS}.
In \cite{netcritic1} Floyd and Paxson, though not addressing physics
approaches directly, note that the while simulations are crucial to
Internet research, the Internet is extraordinarily complicated and
difficult to accurately simulate, especially on large scales. In
particular, they point out three problems: the increasingly
unpredictable behavior of IP over increasingly diverse networks and
applications, the massive and continuously increasing size of the
Internet, and its penchant for changing in many drastic ways over
time. Heterogeneity is the rule not the exception and many
activities such as periodicities are often left out of simulations.
The papers \cite{selfsimilar7, netcriticnew7, netcritic3} address
the physics community more directly pointing out defects in the
theories of critical phenomena and the hub vulnerability of
scale-free networks respectively. In \cite{selfsimilar7}, Willinger
et. al. describe evocative models, which reproduce the observations
using generic models, and explanatory models, whose applicability
are tested by experiment and measurement (what they term ``closing
the loop''). They complain many models, from physicists and some
engineers, are evocative and ignore the research on the particulars
of Internet protocols, function, and traffic that could verify or
refute their model. These models instead act as a ``black box''
which try to come up with an appealing model or theory to match
self-similarity but do not provide any prediction or verification
using real world traffic trace data. For example, the largest
problem in many of the phase transition models is that they
demonstrate self-similar traffic only at critical loads while
Internet traffic measurements show self-similar traffic in both
free-flow and congested regimes and at all levels of flow. In
\cite{netcriticnew7}, Alderson and Willinger further elaborate that
they believe models from statistical mechanics are not applicable to
the Internet which is designed by multiple economic and performance
considerations and not by simple rules of self-organization often
present in models such as preferential attachment.

Barab\'{a}si-Albert preferential attachment model for the growth of
scale-free networks is criticized since many of the highly connected
nodes in the Internet are at the edges near final consumers rather
than in the central parts of the high speed AS network. This valid
criticism though is partially answered by disassortative mixing,
which shows that in non-social networks like the Internet, high
degree nodes are more likely to connect with low-degree nodes and
rules out a core of highly connected, highly vulnerable hubs. Many
subsequent models of scale-free networks have taken this into
account. Finally, \cite{netcritic3,netcriticAMS} criticize the research in
topology that says scale-free networks are vulnerable to attack due
to highly connected hubs, which they once again say is fallacious
because despite power-law degree distributions the most highly
connected hubs are often on the periphery of the Internet and not
along its crucial backbone. They blame much of the confusion on the methodology in the widely cited paper by Faloutsos et. al \cite{faloutsos1}. He et. al \cite{topologyIXP} partially rebut criticisms by studying relationships between customer-providers and peers at Internet exchange points (IXPs). They find the customer provider relationship closely fits a power law though the peer-peer relationships show a Weibull distribution.

Lee in \cite{DFA1} points out both the
aforementioned problems with the critical phenomena models but also
points out the ON/OFF model also has problems because since an
ON/OFF source has a long-tail duration time distribution, you will
have a finite probability of an ON/OFF source as long in duration as
any observation period you make. Lee also criticizes TCP models for
not accounting for similar self-similarity effects in UDP and other stateless
protocols. In \cite{TCPchaos5,TCPchaos8,TCPchaos9,TCPchaos10}, the
possibility is also raised that that TCP can effect the traffic
dynamics in a more market dynamic on shorter time scales (near the
order of multiple RTT) and this is an area of future investigation
and debate.

In the author's opinion, these criticisms of critical phenomena models
are extremely valid and that self-similarity only at critical loads
and lack of real world validation show many of the models in the
previous section are unrealistic or misleading. Given the current
facts and validation against real data, it seems that the current
weight of the evidence for long timescale self-similarity lies with
the application layer explanation based on ON/OFF sources. However,
this does not mean the physics models based on statistical mechanics
are completely useless. Wide scale Internet traffic measurements are
nearly impossible currently and also large-scale theories of traffic
are still relatively undeveloped. Currently almost all real data is
traffic traces over one given link during a given time period which
makes large scale Internet studies very difficult. Large-scale
congestion, traffic correlations, and other complicated phenomena
will probably draw useful lessons from physics models of
self-organization and long-range correlations though more realistic
models are absolutely necessary. The omission of TCP-like congestion
control, except in a few models, must be rectified. If these
criticisms sound a bit harsh, try to put yourself in the shoes of
most network engineers who understand the intricate processes in
detail of how the Internet operates. When shown a model of a 2D
grid, no mention of congestion control, infinite router buffers, and
self-similar traffic only in congested conditions, their incredulity
is understandable. It is aggravated by the fact that almost none of
these papers try to match results with or analyze real traffic
traces.

In defense of the efforts of physicists, however, I believe that
physics started out correctly choosing simplistic topologies and
dynamics scenarios that are both analytically tractable and amenable
to rapid simulation. However, though the earlier work of those like
Takayasu began by looking at traffic traces, this soon disappeared
almost completely in favor of computer simulation. Despite the
obvious shortcomings of explaining self-similar traffic, the
demonstration that large-scale congestion, though on which timescale
is unknown, may be a theory the physics and engineering communities
should take note of for validation or refutation.

With the work in \cite{netcongest8b,netcongest8,netcongest9} showing
large-scale correlations among router traffic in both real and
simulated data, can we really look at the Internet dynamics from
solely the viewpoint of a collection of single traffic traces? The
question is not if the Internet displays large-scale correlations
and self-organization well-known to complexity theory, but how these
large-scale effects play out and if realistic simulations with both
realistic dynamics and topology can predict effects that we have not
yet observed or known how to look for. Much more cross-disciplinary
work is needed in this direction.

Though the analysis of the criticisms above seems like it tries to
be even handed and please everyone while solving nothing, the nature
of the problem is such that the issues regarding the core nature of
Internet traffic cannot be easily resolved. Willinger et. al. are
right in that the loop must be closed and just creating a simulation
that outputs traffic with a Hurst exponent near 0.8 cannot be
considered the final word in the ``cause'' of self-similarity in
Internet traffic. In addition, though it is difficult and near
impossible, large-scale and coordinated traces and models of a
topologically and dynamically correct Internet is the next logical
step in modeling and studying these phenomena.

\section{Other Interesting Phenomena}
\subsection{Flows and fluctuations}
Barab\'{a}si and Argollo de Menezes from the physics community
\cite{fluctuations1} proposed an interesting result when they
announced a relationship between the average volume of the flow and
its dispersion (standard deviation of traffic volume) among nodes in
a network. In particular, they found that accounting for all nodes
in a network, you find the average flux $\langle{f}\rangle$ and
standard deviation $\sigma$ per node are related by the scaling
relationship

\begin{equation}
\label{fluxfluctuation} \sigma \approx \langle{f}\rangle^{\alpha}
\end{equation}

Where $\alpha$ is near either 1 or 1/2 for two types of systems. The
traffic on nodes of a network of Internet routers and on/off state
occurrence of junctions in a microprocessor electronic network had
scaling exponents of 1/2  while visitor traffic to a group of WWW
pages, traffic at a group of highway junctions, and water flow in
different locations in a river network demonstrated scaling
exponents of 1. In two simulations, one based on random walks on a
scale-free network and the other by simulating shortest-path traffic
on a scale-free network, they were able to explain the scaling
exponent of 1/2  as being based on the channeling of traffic through
select nodes and arises from internal or endogenous network
dynamics. The power scaling exponent of 1 on the other hand is shown
to be universal when the amount of traffic is driven by external
forces as well as endogenous dynamics similar to an open system.
They believe the power exponent of 1 is more universal than ½ since
it results from the interplay of endogenous and exogenous pressures.
In a subsequent paper \cite{fluctuations2}, they give a method of
extracting the endogenous and exogenous traffic and propose a
metric, $\eta_{i}$, that defines the predominance of external or
internal influences on traffic dynamics by the equation

\begin{equation}
\label{fluxmetric} \eta_{i} =
\frac{\sigma^{ext}_{i}}{\sigma^{int}_{i}}
\end{equation}

Where $\eta_{i} \gg 1$ indicates an externally driven system while
$\eta_{i} \ll 1$ indicates a systems dominated by internal dynamics.
$\eta_{i}$ can vary on different time scales as \cite{fluctuations3}
showed using trading records from the New York Stock Exchange where
internal dynamics were dominant on the scale of minutes while
external ones were dominant on the hours and days time scales. In
\cite{netroute4b2}, a power law-scaling relationship was also found
via an NNN routing simulation on a scale-free network and scaling
was demonstrated also exhibiting either an exponent of 1/2  or 1.

The generality of the results and the universality of the classes
proposed in these papers has recently been disputed though. Duch and
Arenas \cite{fluctuations4} perform several measurements relating
flow and fluctuations on data from the Abilene Internet backbone and
claim that $\alpha$ varies between 0.71 and 0.86 and not 1/2 as
represented in the first papers. They also propose that this
derivation is due to the original papers disregarding congestion in
networks and show analytically that for short timescales of
measurement, an $\alpha$ of 1/2 is a trivial result but a false
generality once the timescales are extended and other parameters
come into play. They conclude that there is a scaling relationship
but no universality classes as claimed. Meloni et. al.
\cite{fluctuations5} go even further and say that under certain
conditions, power-law scaling between flows and fluctuations should
be abandoned. They conduct a simulation of a random diffusion
process on a network of packets measuring scaling as influenced by
the time window of measurements, the degrees of the nodes flows and
fluctuations are measured on, and the volume of packets in the
network. They produce an analytical result that explains power
scaling behavior between the two quantities only under the
conditions where the noise fluctuations in the system and/or the
time window size are relatively small. Otherwise $\alpha$ tends
towards 1 and does not display power-law scaling. Also, they show
even in networks with power-law scaling, $\alpha$ can scale
differently at 1/2  for low-degree nodes or 1 for high degree nodes
showing that within networks there may be varying scaling depending
on the degrees of the nodes.

Finally, Han et. al. in \cite{fluctuations6} measure $\alpha$ for
the download rates of an Econophysics web database and find an
$\alpha$ varying from 0.6 to 0.89 depending on the length of the
sample time window. They confirm the power law scaling between flux
and fluctuations but do not find any universal exponents. From these
results the research between flows and fluctuations in networks is
still in its earliest stages but holds out much promise for future
progress.

\subsection{Internet worm traffic \& BGP storms}

In \cite{netcongest9}, a simulation by Yuan and Mills was touched on
that aimed to try to predict part of the large-scale impact of a
rapidly spreading Internet worm. Recent increases in the amount and
sophistication of malicious code released on the Internet including
the use of ``zombie'' computers for large-scale DDoS attacks has
demonstrated this is far from just a theoretical exercise. An
increasingly large literature base on the Internet traffic effects
of epidemics has arisen, particularly after the Code Red outbreak in
2001 (which the author had the dubious honor of handling as a
network security administrator at the time). Again, to stay with the
scope of the paper the aspects of Internet worms discussed here will
be tightly limited to effects on traffic, both measured and
predicted, and will not delve into the voluminous theoretical work
of epidemiology on scale-free networks or other topologies
\cite{vespignani1} or much of the new literature with specialized
epidemic models for computer worms and the effect of topology on their spread \cite{epidemic1,epidemic2,epidemic3,epidemic4,epidemic5}. Suffice to say, mathematical results indicate for an ideal epidemic spreading on a scale-free topology, there is no minimum epidemic threshold and that an infection of any size can theoretically spread through the entire network.

The two most studied Internet worms have been the Code Red (start:
July 19, 2001), Nimda (start: September 19, 2001), and SQL Slammer
(start: January 25, 2003). The Slammer, though not holding a
malicious payload, was the fastest spreading worm in history
\cite{slammer}. What has often been found is that the worms not only
cause trouble for the computers they affect, they create large-scale
traffic patterns that can disrupt the normal behavior of entire
networks. Often, a worm spreads by exploiting a vulnerability in
computers and will try to infect random computers by testing an IP
address at random or due to certain rules. With potentially millions
of computers sending out such probes at once it is easy to see how
normal traffic patterns can be seriously disrupted.

In particular, worms have often been the culprit of what could be
termed a large-scale instability in the BGP routing system called a
BGP update storm or BGP storm. In a BGP storm, the normal level of
BGP updates sent to update the router table can rise by several
orders of magnitude and sometimes disrupt traffic \cite{bgpstorm1,
bgpstorm2}. For example, in \cite{bgpstorm1} they describe how
during the Nimda worm normal BGP update traffic of 400
advertisements per minute jumped to 10,000 advertisements per
minute. This is not because the worms infected the routers
themselves but because the worms caused large packet flows which
overwhelmed the router memory and CPU limits and caused them to
crash. These router crashes caused frenzied reorienting of the
Internet router topology. BGP storms are interesting in that both
traffic and topology is rapidly changing. BGP storms may be an
avenue for both physicists and engineers to investigate the
relationship between topology and traffic in a situation when both
are largely in flux.

Yuan and Mills expanded their work from \cite{netcongest9} to a full
paper \cite{ddos} that looks at spatiotemporal correlations between
routers and hosts in several types of large scale DDoS attacks. They
find that DDoS attacks may cause traffic variations at correlated
routers and hosts besides just the target. Because of these large
traffic altering phenomena, certain spectral techniques have been
researched to identify DDoS attacks. Some of these are summarized in
the next section.

\begin{figure}[ht]
\centering
    \includegraphics[height=3.5in, width=3.5in]{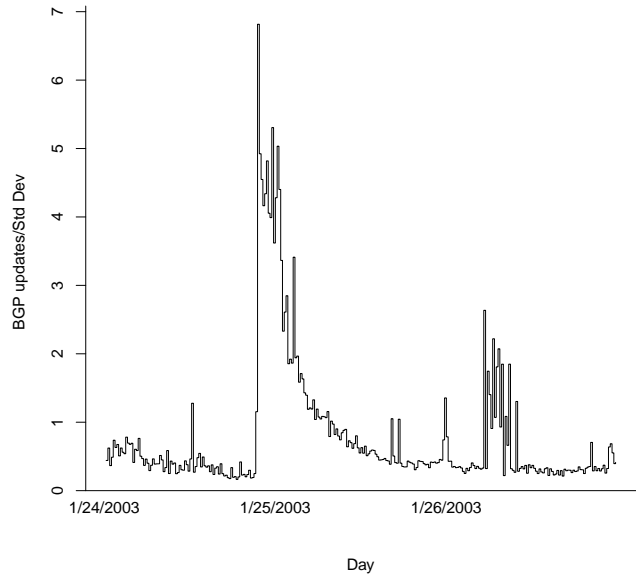}
\caption{A view of a historical BGP storm caused by the onset of the Slammer worm on January 25, 2003 similar to charts in \cite{bgpstorm2}. The chart shows the number of BGP updates divided by the standard deviation of the number of updates over three days, January 24-26, 2003, based on 15 minute intervals. Updates are from routers at the University of Oregon's RouteViews project.} \label{bgpstorm}
\end{figure}

\subsection{Traffic oscillations/periodicities}
Earlier periodic behavior in Internet traffic was casually mentioned
as theoretical assumptions of TCP traffic. Also, one of the
consequences the self-similar nature of traffic is the $1/f$
spectral behavior of the traffic. Beyond these, however, there are a
plethora of traffic periodicities that represent oscillations in
traffic over periods of several scales of magnitude from
milliseconds to weeks. Many of these are well-defined and
classified. Their origin has two possible sources: first, software
or transmission driven periodicities which range on the time scale
of milliseconds, seconds, or in rare cases, hours. Second are user
driven periodicities which range on the time scale of days, weeks,
and possibly longer. This new area of research has been dubbed
\emph{network spectroscopy}\cite{spectrum1}  or \emph{Internet
spectroscopy} and is finding uses in applications such as
identifying traffic sources via traffic periodicity ``fingerprints''
to early detection of denial of service or other hacker attacks by
detecting anomalous oscillations in the traffic spectrum similar to
vibration analysis of faulty machinery.

\begin{figure}[h!]
    \centering
    \includegraphics[height=3.25in, width=6in]{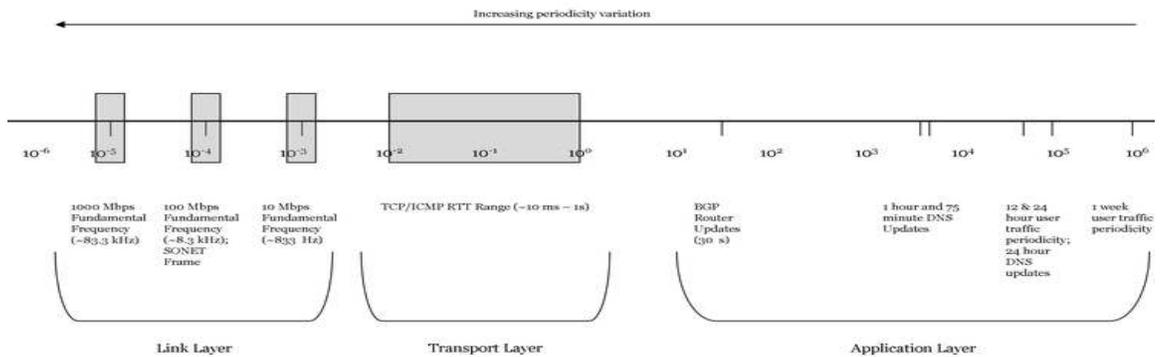}
        \caption{A rough breakdown of the major periodicities in Internet traffic showing the responsible protocols and
        their period in seconds. The periodicities span over 12 orders of magnitude and different protocol layers tend
        to operate on different time scales.}
    \label{periodicities}
\end{figure}

The causes and periods of various known periodicities are summarized
in figure \ref{periodicities}. The values can have a general range
of deviation so the periods are not always exact, but are a good
guide to the major periodicities. User traffic driven periodicities
were the first known and most easily recognized. The first
discovered and most well-known periodicity is the 24 hour diurnal
cycle and its harmonic of 12 hours. These cycles have been
known for decades and reported as early as 1980 and again in 1991 as
well as in many subsequent
studies\cite{spectrum2,spectrum2b,spectrum2c,spectrum2d,spectrum2e,spectrum2f,spectrum3}.
This obviously refers to the 24 hour work-day and its 12-hour second
harmonic as well as activity from around the globe. The other major
periodicity from human behavior is the week with a period of 7 days
\cite{spectrum2c,spectrum2d,spectrum4} and a second harmonic at 3.5
days and barely perceptible third harmonic at 2.3 days. There are
reports as well of seasonal variations in traffic over months
\cite{spectrum5}, but mostly these have not been firmly
characterized. Long period oscillations have been linked to possible
causes of congestion and other network behavior related to network
monitoring \cite{spectrum2e, spectrum2f}. One note is that user
traffic driven periodicities tend to appear in protocols that are
directly used by most end users. The periodicities appear TCP/IP not
UDP/IP and are mainly attributable to activity with the HTTP and
SMTP protocols. They also often do not appear in networks with low
traffic or research aims such as the now defunct 6Bone IPv6 test
network.

The autonomous, non-user driven, periodicities operate mostly at
timescales many orders of magnitude smaller than user behavior. At
the lowest period, and correspondingly highest frequency, are the
periodicities due to the throughput of packet transmission at the
link level. This has been termed the ``fundamental frequency''
\cite{spectrum5} of a link and can be deduced from the equation:

\begin{equation}
\label{linkfreq} f = \frac{T}{s}
\end{equation}

where $T$ is the average throughput of the link and $s$ is the
average packet size at the link level. A quick inspection reveals
this equation is identical to that for the flow rate given by
equation \ref{basiceq}. Indeed, the fundamental frequency is the
rate of packet emission across the link and is the highest frequency
periodicity possible. The theoretical maximum fundamental frequency
is given by

\begin{equation}
\label{linkfreqmax} f_{max}  = \frac{B}{MTU}
\end{equation}

where B is the bandwidth of the link and the packet size is the MTU
packet size. Therefore for 1 Gigabit, 100 Mbps, and 10 Mbps Ethernet
links with MTU sizes of 1500 bytes, the theoretical maximum
fundamental frequencies are 83.3 kHz, 8.3 kHz, and 833 Hz
respectively.

The usual measured fundamental frequencies via power spectrum
diagrams are lower than the theoretical fundamental frequencies due
to lower throughput. The fundamental frequency also generally
displays harmonics as well \cite{spectrum5}.

Broido, et. al. \cite{spectrum6} believe there are thousands of
periodic processes in the Internet. Among other prominent recognized
periodicities are BGP router table updates sent every 30 seconds,
SONET frames transmitted every 125$\mu$s, DNS updates transmitted
with periods of 75 minutes, 1 hour, and 24 hours due to default
settings in Windows 2000 and XP DNS software\cite{spectrum1}, and in
TCP flows ACK packets at a frequency of 1/RTT \cite{spectrum6,
spectrum8, spectrum9} with RTTs usually ranging from 10ms to 1s.

The main practical applications being researched for network
spectroscopy are inferring network path characteristics such as
bandwidth, digital fingerprinting of link transmissions, and
detecting malicious attack traffic by changes in the frequency
domain of the transmission signal. In \cite{spectrum10, spectrum10b}, the authors use analysis of the distribution of packet interarrival times to infer
congestion and bottlenecks on network paths upstream. In
\cite{spectrum9, spectrum11, spectrum12,spectrum13, spectrum14}
various measures of packet arrival distributions, particularly in
the frequency domain, are being tested to recognize and analyze
distributed denial of service or other malicious attacks against
computer networks. Inspecting the frequency domain of a signal can
also reveal the fingerprints of the various link level technologies
used along the route of the signal as is done in \cite{spectrum6,
spectrum15}.

\subsection{Biological/ecological models and Internet traffic}

Comparisons of the Internet to biological or ecological systems are
legion and range from the theoretically precise to philosophical
speculations in both popular fiction such as William Gibson's
Neuromancer and Masamune Shirow's Ghost in the Shell
\cite{book1,book2} as well as in the opinions of some researchers
such as Vernor Vinge's ``Singularity''\cite{singularity}.

The focus here is on scientific papers which have used mathematical
models, biological or ecological, to describe functions of the
Internet or compare certain functions to physical systems. The
growth of the Internet's nodes in terms of a birth/death process is
covered in \cite{review7}. In \cite{ANS} Fukuda, Nunes-Amaral, and
Stanley use several statistical analyses to show a striking
similarity between variations in daily Internet active connections
in a data trace and statistics on heartbeat intervals. By both
separating both non-stationary time series into stationary segments
and using DFA, they show that the magnitudes of activity for night
time (non-congested) Internet connections and healthy heartbeats are
statistically very similar. Likewise day time (congested) Internet
connections and diseased heartbeat intervals are also similar in
their fluctuations. They propose that a general nonlinear systems
explanation underlies both systems and given that the heart rate is
controlled by the autonomic nervous system, understandings of
Internet functions and properties could be used to study the
autonomic nervous system as well.

Several authors have also used ecological interaction models such as
mathematical models of competition and mutualism to study
interaction between web sites and search engines. In \cite{comp1,
comp2, comp3}, competition and cooperation between web sites are
analyzed using the n-competitor Lotka-Volterra differential
equations from ecology. The steady state of ``winner takes all'' or
multiple participants is extracted from stability criterion and
compared to actual market competition. In \cite{comp4}, another
model which includes a cooperation effect is introduced to study the
same dynamics. The interesting analogy between search engines and
websites as a mutualistic relationship is introduced in
\cite{mutualism}. The postulated mutualism is obligate for the
search engine and facultative for the websites. This is similar to
the sea anemone and hermit crab or mycorrhiza and plant mutualisms
in nature. They show that strong mutual support for web sites by
search engines and vice versa offers the best opportunity for
long-term sustainability and growth.

The last paper covered in this section is a recent publication which
draws similarities between the energy use and scaling of information
networks and metabolic scaling phenomena such as Kleiber's Law, the
3/4  power law scaling of organism mass and metabolism
\cite{infonet}. Though the bulk of the paper is comparing the
circuitry density and area for electronic circuits microprocessors,
they derive, with a limited set of data points, a scaling
relationship between the total processing power of hosts on the
Internet and Internet backbone bandwidth with a scaling exponent of
about 2/3. Future research in this direction, especially if a valid
scaling law relating topology and dynamics is discovered, will
surely be very fruitful.

\begin{table*}[!t] \vspace{1.5ex}
\scriptsize
\centering 
\begin{tabular}{p{0.7 in} p{3.5 in}}
\hline
Name&Data\\
\hline
CAIDA&Probably largest and most comprehensive repository of all types of Internet data and research. Hosted by UC San Diego\\
Crawdad&Collection of wireless data traces made at Dartmouth\\
Datapository&Collection of topology data, mostly based on BGP\\
Dimes&A project that seeks to more accurately measure Internet topology by applying traceroute and ping from a collection of different users with downloaded software around the world in a model similar to SETI@Home\\
MOME&Large collection of traffic traces conducted by European based researchers and collaborations\\
NLANR&Older traffic trace project; now mostly housed at CAIDA\\
ns-2&Commonly used network traffic simulator in the network
engineering community\\
PingER&Stanford project to monitor Ping response in IPv4 and IPv6 across the Internet\\
PREDICT&A collection of datasets of Internet topology and traffic tailored towards predicting and defending against cyberattacks. Must apply for data access\\
RouteViews&U of Oregon's database on Internet routing tables and BGP data\\
tcpdump&Main program used to collect traffic for analysis; used in many packet sniffing programs\\
UMass Trace Repository&Collection of publicly available traffic traces conducted at the University of Massachusetts Amherst\\
WAND&Collection of traces done by the University of Waikato, New Zealand\\
WIDE MAWI&Japan's WIDE Project traffic trace archive; data source of many graphs in this paper\\

\hline
\end{tabular}
\label{datasources} \caption{Common Internet traffic data sources \&
software.}
\end{table*}
\section{Conclusion}

With every passing year, research is making us more and more aware
of the complex dynamics and interplay of factors on the Internet.
Though many may haphazardly use the terms self-organization,
emergence, or power law this review has hopefully laid out the
concrete facts about what is known clearly about Internet traffic,
what is less clear, and where many new paths can be beaten. Unlike
most systems which are amenable to constant analysis over long time
periods, the Internet is ever changing. What we understand today may
not completely apply several years from now. In addition, our
knowledge of long range correlations and dynamics among multiple
sites and links is still in its infancy. Large-scale congestion is the only possible large-scale property which has been studied in any detail and it
remains to be seen if it is the only one that exists. There is much
room for speculation on these matters without being irresponsibly
fanciful.

At the core, these issues are more than academic since the long-term
efficiency and stability will require us to understand the Internet
and its traffic well enough to optimize it for the ends of users.
Advances in understanding the Internet are also enabled and
constrained about our knowledge of nonlinear dynamics and complex
systems in general. As more themes and discoveries about these
systems emerge, they will doubtlessly provide us with more tools
with which to investigate the Internet and uncover more of the story
behind its dynamics. Finally, as mentioned earlier, it is essential
for more cross-disciplinary cooperation to take place in order to
accelerate our understanding of Internet phenomena. The two groups
have cooperated in some areas and are hardly irreconcilable.
Combining both toolkits can definitely bring forth some more
surprising and rewarding results.

Finally, though this paper has been heavy on esoteric technical
aspects of the Internet, we must not lose sight of the whole, as the
poet Walt Whitman once wonderfully wrote \cite{leaves},
\begin{verse}
When I heard the learn'd astronomer; \\
When the proofs, the figures, were ranged in columns before me; \\
When I was shown the charts and the diagrams, to add, divide, and measure them; \\
When I, sitting, heard the astronomer, where he lectured with much applause in the \\ lecture-room,  \\
How soon, unaccountable, I became tired and sick;            \\
Till rising and gliding out, I wander'd off by myself,   \\
In the mystical moist night-air, and from time to time,  \\
Look'd up in perfect silence at the stars. \\
\end{verse}

For everything said about self-similarity, phase transitions, and
related matter we must never lose sight of the Internet as the
wonderful invention it has been in its cultural, economic, and
technological aspects uniting those from around the world. Even if
in only a small part, this should animate and encourage our research
into the future.


\begin{thebibliography} {1}
\bibitem{netcongest2x}Abe, S. \& Suzuki, N., Omori's law in the Internet traffic, \emph{Europhys. Lett.} \textbf{61} (2003), 852-855.
\bibitem{multifractal2b}Abry, P. \& Veitch, D., Wavelet Analysis of Long-Range Dependence Traffic, \emph{IEEE Transactions on Information Theory}, \textbf{44}:1 (1998), pp.2-15.
\bibitem{selfsimilar10}Abry, P., Baraniuk, R., Flandrin, P., Riedi, R., \& Veitch, D., Multiscale Network Traffic Analysis, Modeling, and Inference Using Wavelets, Multifractals, and Cascades, \emph{IEEE Signal Processing Magazine}, \textbf{19}:3 (2002), 28-46.
\bibitem{logscalegood1}Abry, P., Flandrin, P., Taqqu, M.S. \& Veitch, D., Wavelets for the analysis, estimation, and synthesis
of scaling data, In \emph{Self-similar network traffic and performance evaluation} edited by Park, K. \& Willinger, W. (John Wiley \& Sons, New York, 2000) pp. 39-88.
\bibitem{logscalegood2}Abry, P., Flandrin, P., Taqqu, M.S. \& Veitch, D. Self-similarity and long-range dependence through
the wavelet lens, In \emph{Theory and Applications of Long-Range Dependence} edited by Doukan, P., Taqqu, M.S., Oppenheim, G. (Birkh¨auser, Boston, 2003) pp. 527-556.
\bibitem{netroute4}Adamic, L., Lukose, R., Puniyani, A., \& Huberman, B. Search in power-law networks, \emph{Phys. Rev. E}, \textbf{64} (2001), 046135.
\bibitem{review1}Albert, R. \& Barab\'{a}si, A.L., Statistical mechanics of complex networks, \emph{Rev. Mod. Phys.}, \textbf{74} (2002), 47-97.
\bibitem{netcriticnew7}Alderson, D.L. \& Willinger, W., A contrasting look at self-organization in the Internet and next-generation communication
networks, \emph{IEEE Communications Magazine}, \textbf{43}:7, (2005)
94-100 .
\bibitem{netcriticnew2b}Alderson, D.L., Li, L., Willinger, W., \& Doyle, J. Understanding Internet topology: Principles, models, and
validation, \emph{IEEE/ACM Transactions on Networking},
\textbf{13}:6, (2005) 1205-1218.
\bibitem{netcriticnew5}Alderson, D.L., Chang, H., Roughan, M., Uhlig, S. \&
Willinger, W., The many facets of Internet topology and traffic,
\emph{Networks and Heterogenous Media}, \textbf{1}:4, (2006)
569-600.
\bibitem{netcongest2c}Arenas, A., Di\`{a}z-Guilera, A. \& Guimer\'{a}, R., Communication in Networks with Hierarchical Branching, \emph{Phys. Rev. Lett.}, \textbf{86} (2001), 3196-3199.
\bibitem{fluctuations1}Argollo de Menezes, M, \& Barab\'{a}si, A.L., Fluctuations in Network Dynamics, \emph{Phys. Rev. Lett.}, \textbf{92} (2004), 028701.
\bibitem{fluctuations2}Argollo de Menezes, M. \& Barab\'{a}si, A.L., Separating Internal and External Dynamics of Complex Systems, \emph{Phys. Rev. Lett.}, \textbf{93}
(2004), 068701.
\bibitem{selfsimilarcause4}Baker, M.G., Hartman, J.H., Kupfer, M.D., Shirriff, K.W., \& Ousterhout, J.K., Measurements of a distributed file system , in \emph{Proceedings of the 13th ACM Symposium on Operating System Principles},  (Pacific Grove, CA ,1991), pp. 198-212.
\bibitem{Barabasi1} Albert, R., Jeong, H., \& Barab\'{a}si, A.L., Internet — diameter of the World-Wide Web, \emph{Nature}, \textbf{401} (1999), 130131 .
\bibitem{dfacritic}Bardet, J.M. \& Kammoun, I., Asymptotic properties of the detrended fluctuation analysis of long-range-dependent processes, \emph{IEEE Transactions on
Information Theory}, \textbf{54}:5, (2008) 2041-2052.
\bibitem{netcongest8b}Barth\'{e}lemy, M., Gondran, B., Guichard, E., Large scale cross-correlations in Internet traffic, \emph{Phys. Rev. E}, \textbf{66}  (2002), 056110.
\bibitem{elephant6}Barth\'{e}lemy, M., Gondran, B., \& Guichard, E., Spatial structure of the internet traffic,  \emph{Physica A}, \textbf{319}:1 (2003), 633-642.
\bibitem{protocol7}Basher, N., Mahanti, A., Williamson, C. \& Arlitt, M., A comparative analysis of web and peer-to-peer traffic, \emph{Proceedings of the 17th International Conference on the
World Wide Web} (Beijing, China, 2008), pp. 287-296.
\bibitem{review4}Boccaletti, S., Latora, V., Moreno, Y., Chavez, M., \& Hwang, D.U., Complex networks: Structure and dynamics, \emph{Phys. Repp.}, \textbf{424}:4-5 (2006), 175-308.
\bibitem{spectrum8}Broido, A., Invariance of Internet RTT spectrum, in \emph{Proceedings of ISMA Conference}, (San Diego, CA, 2002).
\bibitem{spectrum1}Broido, A., Nemeth, E., \& Claffy, K.C., Spectroscopy of DNS Update Traffic, \emph{ACM SIGMETRICS 2003}, \textbf{31}:1 (2003) 320-321.
\bibitem{spectrum6}Broido, A., R King, Nemeth, E., Claffy, K.C., Radon Spectroscopy of Packet Delay, in \emph{Proceedings of the IEEE High-Speed Networking Workshop 2003}, (San Diego, CA 2003).
\bibitem{dragonfly}Brownlee, N. \& Claffy, K.C., Understanding Internet traffic streams: dragonflies and tortoises, \emph{IEEE Communications Magazine}, \textbf{40}:10 (2002) 110-117.
\bibitem{spectrum4}Burgess, M., Haugerud, H., Straumsnes, S., \& Reitan, T., Measuring system normality, \emph{ACM Trans. on ComSys.}, \textbf{20}:2 (2002) 125-160.
\bibitem{protocol1}C\'{a}ceres, R., Measurements of wide-area Internet traffic University of California Berkeley Technical Report, CSD-89-550 (1989).
\bibitem{protocol2}C\'{a}ceres, R., Danzig, P., Jamin, S., \& Mitzel, D., Characteristics of wide-area TCP/IP conversations,  \emph{Proceedings of SIGCOMM '91}, \textbf{21}:4 (Zurich, Switzerland, 1991), pp. 101-112.
\bibitem{netcongest4b}Campos, I., Taranc\'{o}n, A., Cl\'{e}rot, F., \&  Fern\'{a}ndez, L.A., Thermal and repulsive traffic flow, \emph{Phys. Rev. E}, \textbf{52} (1995) 5946-5954.
\bibitem{HOTmodel}Chang, H.S., Jamin, S., \& Willinger, W., Internet connectivity at the AS-level: an optimization-driven modeling approach, \emph{Proceedings of the ACM SIGCOMM workshop on Models, methods and tools for reproducible network research}, (Karlsruhe, 2003) pp. 33-46.
\bibitem{spectrum11}Chen, Y. \& Hwang, K., Collaborative detection and filtering of shrew DDoS attacks using spectral analysis, \emph{Journal of Parallel and Distributed Computing,}
\textbf{66}:9 (2006) 1137-1151.
\bibitem{netroute5a}Chen, Z.Y. \& Wang, X.F., Effects of network structure and routing strategy on network capacity, \emph{Phys. Rev. E}, \textbf{73} (2006) 036107.
\bibitem{spectrum9}Cheng, C.M., Kung, H.T., \& Tan, K.S., Use of spectral analysis in defense against DoS attacks, in \emph{Proceedings of IEEE GLOBECOM '02}, \textbf{3} (Taipei, Taiwan, 2002) pp. 2143-2148.
\bibitem{vehicle3}Chowdhury, D., Santen, L., \& Schadschneider, A., Statistical physics of vehicular traffic and some related systems, \emph{Phys. Repp.}, \textbf{329}:4-6 (2000) 199-329.
\bibitem{packetsize4}Claffy, K.C. \& McCreary, S., Trends in Wide Area IP Traffic Patterns: A View from Ames Internet Exchange, \emph{Proceedings of the 13th ITC Specialist Seminar on Internet Traffic Measurement and Modeling} (Monterey, CA, 2000).
\bibitem{protocol3}Claffy, K.C. \& Polyzos, G.C., Traffic characteristics of the T1 NSFNET backbone,  \emph{Proceedings of INFOCOM '93} (San Francisco, CA, 1993) pp. 885-892.
\bibitem{packetflowdef2}Claffy, K.C., Braun, H.W., \& Polyzos, G.C., A parameterizable methodology for Internet traffic flow profiling,, \emph{IEEE Journal on Selected Areas in Communication}, \textbf{13} (1995) 1481-1494.
\bibitem{powerlawsense1}Clauset, A., Shalizi, C.R., \& Newman, M.E.J., Power-law distributions in empirical data, to
appear \emph{SIAM Review}, preprint arXiv:0706.1062 (2009).
\bibitem{hurstguide}Clegg, R.G., A practical guide to measuring the Hurst parameter, \emph{International Journal of Simulation: Systems, Science \& Technology}, \textbf{7}:2 (2006) 3-14.
\bibitem{spectrum15}Coates, M., Hero, A., Nowak, R. \& Yu, B., Internet tomography, \emph{IEEE Signal Processing Magazine}, \textbf{19}:3 (2002) 47-65.
\bibitem{bgpstorm1}Cowie, J., Ogielski, A.T., Premore, B.J., \& Yuan, Y.G., Internet worms and global routing instabilities, in \emph{Proceedings of SPIE}, \textbf{4868} (San Jose, CA, 2002) pp. 195-199.
\bibitem{selfsimilarcause3b}Crovella, M.E. \& Bestavros, A., Self-similarity in World Wide Web traffic: evidence and possible causes , \emph{IEEE ACM Transactions on Networking}, \textbf{5}:6 (1997) 835-846.
\bibitem{cascade3}Crucitti, P., Latora, V., \& Marchiori, M., Model for cascading failures in complex networks, \emph{Phys. Rev. E}, \textbf{69} (2004) 045104.
\bibitem{netcongest1}Csabai, I.J., 1/f noise in computer network traffic, \emph{Phys. A: Math. Gen.}, \textbf{27} (1994) 417-421.
\bibitem{vehicle2}Daganzo, C.F., Cassidy, M.J. \& Bertini, R.L. Causes and Effects of Phase Transitions in Highway Traffic University of California, Berkeley Institute of Transportation Studies Research Report UCB-ITS-RR-97-8, (1997).
\bibitem{review3}Dorogovtsev, S.N. \& Mendes, J.F.F., Evolution of networks, \emph{Adv. in Phys.}, \textbf{51}:4 (2002) 1079-1187.
\bibitem{packetsize3}Dovrolis, C., Ramanathan, P., \& Moore, D., What do packet dispersion techniques measure?, \emph{Proceedings of INFOCOM 2001},  \textbf{2} (Anchorage, AL, 2001) 905-914.
\bibitem{selfsimilarcause5}Downey, A.B., The structural cause of file size distributions, in \emph{Proceedings of the 9th IEEE/MASCOTS (Modeling, Analysis and Simulation of Computer and Telecommunication Systems)}(Cincinnati, OH, 2001) pp. 361-370.
\bibitem{netcritic3}Doyle, J.C., Alderson, D.L., Li, L., Low, S., Roughan, M., Shalunov, S., Tanaka, R., \& Willinger, W., The `robust yet fragile' nature of the Internet, \emph{Proc. Natl. Acad. of Sci.}, \textbf{102}:41 (2005) 14497-14502.
\bibitem{fluctuations4}Duch, J. \& Arenas, A., Scaling of Fluctuations in Traffic on Complex Networks, \emph{Phys. Rev. Lett.}, \textbf{96} (2006) 218702.
\bibitem{netroute1}Echenique, P., G\'{o}mez-Garde\~{n}as, J., \& Moreno, Y., Improved routing strategies for Internet traffic delivery, \emph{Phys. Rev. E}, \textbf{70} (2004) 056105.
\bibitem{netroute2}Echenique, P., G\'{o}mez-Garde\~{n}as, J., \& Moreno, Y., Dynamics of jamming transitions in complex networks, \emph{Europhys. Lett.}, \textbf{71} (2005) 325.
\bibitem{fluctuations3}Eisler, Z., Kert\'{e}sz, J., Yook, S.H., \& Barab\'{a}si, A.L., Multiscaling and non-universality in fluctuations of driven complex
systems, \emph{Europhys. Lett.}, \textbf{69} (2005) 664-670.
\bibitem{TCPthrough5}El Khayat, I., Geurts, P., \& Leduc, G., On the Accuracy of Analytical Models of TCP Throughput in \emph{Networking Technologies, Services, and Protocols; Performance of Computer and Communication Networks; Mobile and Wireless Communications Systems}, (Springer, Berlin, 2006) pp. 488.
\bibitem{protocol8}Erman, J., Mahanti, A., Arlitt, M., \& Williamson, C., Identifying and discriminating between web and peer-to-peer traffic in the network core, \emph{Proceedings of the 16th International Conference on World Wide Web} (Banff, Alberta, 2007) pp. 883-892.
\bibitem{selfsimilar8}Erramilli, A., Roughan, M., Veitch, D., \& Willinger, W., Self-similar traffic and network dynamics, \emph{Proc. of the IEEE}, \textbf{90} (2002) 800-819.
\bibitem{elephant1}Estan, C. \& Varghese, G., New directions in traffic measurement and accounting: Focusing on the elephants, ignoring the mice, \emph{ACM Trans. on ComSys.}, \textbf{21}:3 (2003) 270-313.
\bibitem{faloutsos1}Faloutsos, M., Faloutsos, P., \& Faloutsos, C., On power-law relationships of the Internet topology \emph{ACM SIGCOMM ComComm. Rev.}, \textbf{29} (1999) 251-262.
\bibitem{elephant2}Fang, W. \& Peterson, L., Inter-AS traffic patterns and their implications, \emph{Proceedings of IEEE GLOBECOM '99} \textbf{3} (Rio de Janeiro, Brazil, 1999) pp. 1859-1868.
\bibitem{hops}Fei, A.G., Pei, G.Y., Liu, R., \& Zhang, L.X., Measurements on Delay and Hop-Count of the Internet, in \emph{Proceedings of IEEE GLOBECOM '98-Internet Mini-Conf.}, (Sydney, Australia, 1998).
\bibitem{TCPchaos2}Fekete, A. \& Vattay, G., Self-similarity in bottleneck buffers, in the \emph{Proceedings of IEEE GLOBECOM '01}, \textbf{3}
(San Antonio, TX, 2001) pp.1867-1871.
\bibitem{multifractal3}Feldmann, A., Gilbert, A.C., Willinger, W., \& TG Kurtz, The changing nature of network traffic: scaling phenomena, \emph{ACM SIGCOMM ComComm. Rev.}, \textbf{28}:2 (1998) 5-29.
\bibitem{multifractal2}Feldmann, A., Gilbert, A.C., \& Willinger, W., Data networks as cascades: investigating the multifractal nature of Internet WAN traffic, \emph{ACM SIGCOMM ComComm. Rev.}, \textbf{28}:4 (1998) 42-55.
\bibitem{TCPchaos8}Figueiredo, D., Liu, B., Misra, V., \& Towsley, D., On the autocorrelation structure of TCP traffic, \emph{Computer Networks}, \textbf{40}:3 (2002) 339-361.
\bibitem{TCPchaos9}Figueiredo, D., Liu, B., Feldmann, A., Misra, V., Towsley, D., \& Willinger, W., On TCP and self-similar traffic, \emph{Performance Evaluation}, \textbf{61}:2-3 (2005) 129-141.
\bibitem{TCPthrough1}Floyd, S., Connections with multiple congested gateways in packet-switched networks part 1: one-way traffic, \emph{ACM SIGCOMM ComComm. Rev.}, \textbf{21}:5 (1991) 30-47.
\bibitem{TCPred}Floyd, S. \& Jacobson, V., Random Early Detection gateways for Congestion Avoidance, \emph{IEEE/ACM Transactions on Networking}, \textbf{1}: 4 (1993) 397-413.
\bibitem{netcritic1}Floyd, S., \& Paxson, V., Difficulties in simulating the internet, \emph{IEEE/ACM Transactions on Networking}, \textbf{9}:4 (2001) 392-403.
\bibitem{protocol6}Fomenkov, M., Keys, K., Moore, D., \& Claffy, K.C., Longitudinal study of Internet traffic in 1998-2003,  \emph{Proceedings of the Winter International Symposium on
Information and Communication Technologies} (Dublin, Ireland, 2004)
pp. 1-6.
\bibitem{spectrum3}Fowler, H.J. \& Leland, W.E., Local area network characteristics, with implications for broadband network congestion management, \emph{IEEE Journal on Selected Areas in Communications}, \textbf{9}:7 (1991) 1139-1149.
\bibitem{protocol5}Fraleigh, C., Moon, S., Lyles, B., Cotton, C., Khan, M., Moll, D., Rockell, R., Seely, T., \& Diot, C., Packet-level traffic measurements from the Sprint IP backbone, \emph{IEEE Network}, \textbf{17}:6 (2003) 6-16.
\bibitem{protocol4b}Frazer, K.D., NSFNET: A Partnership for High-Speed Networking, Final Report 1987-1995, Merit Network, Inc. (1995).
\bibitem{netcongest5}Fuk\'{s}, H. \& Lawniczak, A., Performance of data networks with random links, \emph{Mathematics and Computers in Simulation},
\textbf{51}:1 (1999) 101-117.
\bibitem{ANS}Fukuda, K., Numes Amaral, L.A., \&  Stanley, H.E., Similarities between communication dynamics in the Internet and the autonomic nervous system, \emph{Europhys. Lett.}, \textbf{62} (2003) 189-195.
\bibitem{vehicle4}G\'{a}bor, S. \& Csabai, I.J., The analogies of highway and computer network traffic, \emph{Physica A}, \textbf{307}:3-4 (2002) 516-526.
\bibitem{epidemic1}Ganesh, A., Massoulie, L., Towsley, D., The effect of network topology on the spread of epidemics, in \emph{Proceedings of INFOCOM 2005. 24th Annual Joint Conference of the IEEE Computer and Communications Societies}, \textbf{2} (Miami, 2005) 1455-1466.
\bibitem{multifractal5b}Gao, J.B. \& Rubin, I., Multiplicative multifractal modeling of long-range-dependent network traffic, \emph{Intl. J. of Comm. Sys.}, \textbf{14}:8 (2001) 783-801.
\bibitem{book1}Gibson, W., \emph{Neuromancer} (Ace Books, New York, 1984).
\bibitem{multifractal4}Gilbert, A.C., Willinger, W., \& Feldmann, A. Scaling analysis of conservative cascades, with applications to network traffic,  \emph{IEEE Transactions on Information Theory}, \textbf{45}:3 (1999) 971-991.
\bibitem{selfsimilarcause5b}Gong, W.B., Liu, Y., Misra, V., \& Towsley, D., On the tails of Web file size distributions, in \emph{Proceedings of the 39th Allerton Conference on Communication, Control, and Computing}, (Monticello, Illinois, 2001).
\bibitem{netroute6}Guimer\`{a}, R., Arenas, A., D\'{i}az-Guilera, A., \& Giralt, F., Dynamical properties of model communication networks, \emph{Phys. Rev. E}, \textbf{66} (2002) 026704.
\bibitem{TCPchaos5}Guo, L., Crovella, M., \& Matta, I., How does TCP generate pseudo self-similarity, in \emph{Proceedings of the 9th IEEE/MASCOTS (Modeling, Analysis and Simulation of Computer and Telecommunication Systems)} (Cincinnati, OH, 2001) pp. 215-223.
\bibitem{TCPchaos4}H\'{a}ga, P., Pollner, P., Simon, G., Csabai, I.J., \& Vattay, G., Self-generated self-similar traffic, \emph{Nonlinear Phenomena in Complex Systems}, \textbf{6}:4 (2003) 814-823.
\bibitem{fluctuations6}Han, D.D., Liu, J.G., \& Ma, Y.G., Fluctuation of the Download Network, \emph{Chin. Phys. Lett.}, \textbf{25} (2008) 765-768.
\bibitem{spectrum5}He, X., Papadopoulos, C., Heidemann, J., \& Hussain, A., Spectral Characteristics of Saturated Links, University of Southern California Technical Report, USC-CSD-TR-827 (2004).
\bibitem{spectrum10b}He, X., Papadopoulos, C., Heidemann, J., Mitra, U., Riaz, U., \&  Hussain, A.,  Spectral Analysis of Bottleneck Traffic, University of Southern California Technical Report, USC/CS Technical Report 05-853 (2005).
\bibitem{topologyIXP}He,Y., Siganos, G., Faloutsos, M., \& Krishnamurthy, S., ``A Systematic Framework for Unearthing the Missing Links: Measurements and Impact'' in \emph{4th USENIX Symposium on Networked Systems Design \& Implementation}, (Cambridge, MA, 2007), pp. 187-200.
\bibitem{netroute5also2}Hu, M.B., Wang, W.X., Jiang, R., Wu, Q.S., \& Wu, Y.H., Phase transition and hysteresis in scale-free network traffic, \emph{Phys. Rev. E}, \textbf{75} (2007) 036102.
\bibitem{netroute5also3} Hu, M.B.,, Wang, W.X., Jiang, R., Wu, Q.S., \& Wu, Y.H., The effect of bandwidth in scale-free network traffic,  \emph{Europhys. Lett.}, \textbf{79} (2007) 14003.
\bibitem{spectrum13} Hussain, A., Heidemann, J., \& Papadopolous, C., A framework for classifying denial of service attacks, in \emph{Proceedings of the 2003 Conference on Applications, Technologies, Architectures, and Protocols for Computer Communications} (Karlsruhe, Germany, 2003) pp. 99-110.
\bibitem{spectrum12}Hussain, A., Heidemann, J., \& Papadopolous, C., Identification of Repeated Attacks Using Network Traffic Forensics, USC/ISI Technical Report ISI-TR-2003-577b (2004).
\bibitem{ID2}Iliofotou, M. et. al., Graph-based P2P traffic classification at the Internet backbone, \emph{Proceedings of the 28th IEEE international conference on Computer Communications Workshops} (Rio de Janeiro, 2009) pp. 37-42. 
\bibitem{packetflowdef1}Jain, R. \& Routhier, S., Packet trains: measurements and a new model for computer network traffic,  \emph{IEEE Journal on Selected Areas in Communication}, \textbf{4}:6 (1986) 986-995.
\bibitem{wavelet3}Kaiser, G., \emph{A Friendly Guide to Wavelets}, (Springer, Berlin, 1994).
\bibitem{selfsimilar6}Karagiannis, T., Molle, M., \& Faloutsos, M., Long-Range Dependence: Ten Years of Internet Traffic Modeling, \emph{IEEE Internet Computing}, \textbf{8}:5 (2004) 57-64.
\bibitem{ID1}Karagiannis, T., Papagiannaki, K., \& Faloutsos, M., BLINC: multilevel traffic classification in the dark,\emph{ACM/ SIGCOMM ComComm Review}, \textbf{35}:4 (2004) 229-240 
\bibitem{ID3}Karagiannis, T., Broido, A., Faloutsos, M. \& Claffy, K.C., Transport layer identification of P2P traffic, \emph{Proceedings of the 4th ACM SIGCOMM conference on Internet measurement} (Taromina, 2004) pp. 121-134.
\bibitem{spectrum10}Katabi, D. \& Blake, C., Inferring Congestion Sharing and Path Characteristics from Packet Interarrival Times, MIT Technical Report, MIT-LCSTR-828 (2001).
\bibitem{vehicle1}Kerner, B.S., \& Rehborn, H., Experimental Properties of Phase Transitions in Traffic Flow, \emph{Phys. Rev. Lett.}, \textbf{79} (1997) 4030-4033.
\bibitem{packetsize1}Khalil, K.M., Luc, K.Q., \& Wilson, D.V., LAN traffic analysis and workload characterization, \emph{Proceedings of the 15th Conference on Local Computer
Networks} (Minneapolis, MN, 1990) pp. 112-122.
\bibitem{ID5}Kim, H. et. al., Internet traffic classification demystified: myths, caveats, and the best practices, \emph{Proceedings of the 2008 ACM CoNEXT Conference}, (Madrid, 2008) Article 11. 
\bibitem{epidemic3}Kim, J.H., Radhakrishnan, S., Dhall, S.K., Measurement and analysis of worm propagation on Internet network topology, in \emph{Proceedings of the 13th International Conference on Computer Communications and Networks} (Chicago, 2004) pp. 495-500.
\bibitem{caidameet}Krioukov, D., Chung, F., Claffy, K.C., Fomenkov, M., Vespignani, A., \& Willinger, W., The Workshop on
Internet Topology (WIT) Report, \emph{ACM SIGCOMM ComComm. Rev},
\textbf{37}:1 69-73 (2007).
\bibitem{netcriticnew3}Krishnamurthy, B. \& Willinger, W., What are our standards for validation of
measurement-based networking research?, \emph{Performance Evaluation
Review}, \textbf{36}:2, 64-69 (2008).
\bibitem{packetsize5}Kushida, T., An empirical study of the characteristics of Internet traffic,  \emph{ComComm.}, \textbf{22}:17 (1999) 1607-1618.
\bibitem{Brownian}L\'{e}vy-V\'{e}hel, J. \& Riedi, R., Fractional Brownian motion and data traffic modelling, in \emph{Fractals in Engineering: New Trends in Theory and Applications}, edited by J L\'{e}vy-V\'{e}hel \& E Lutton, (Springer-Verlag, Berlin, 1997) pp. 185-202.
\bibitem{comp2}L\'{o}pez, L. \& Sanju\'{a}n, M., Defining strategies to win in the Internet market, \emph{Physica A}, \textbf{301}:4 (2001) 512-534.
\bibitem{comp3}L\'{o}pez, L., Almendral, J.A., \& Sanju\'{a}n, M., Complex networks and the WWW market, \emph{Physica A}, \textbf{324}:4 (2003) 754-758.
\bibitem{spectrum2b}Lakhina, A., Papagiannaki, K., Crovella, M.E., Diot, C., Kolaczyk, E., \& Taft, N., Structural analysis of network traffic flows, \emph{Performance Evaluation} Review, \textbf{32}:1 (2004) 61-72.
\bibitem{TCPthrough2}Lakshman, T.V. \& Madhow, U., The performance of TCP/IP for networks with high bandwidth-delay products and random loss, \emph{IEEE/ACM Transactions on Networks}, \textbf{5}:3 (1997) 336-350.
\bibitem{elephant3}Lan, K. \& Heidemann, J., On the correlation of Internet flow characteristic, University of Southern California Information Sciences Institute Technical Report,  ISI-TR-574, USC/ISI, (2003).
\bibitem{elephant0}Lan, K. \& Heidemann, J., A measurement study of correlations of Internet flow characteristics, \emph{Computer Networks}, \textbf{50}:1 (2006) 46-62.
\bibitem{netroute5b}Lawniczak, A.T. \& Tang, X.W., Network traffic behaviour near phase transition point, \emph{Eur. Phys. J. B}, \textbf{50}:2 (2006) 231-236.
\bibitem{netroute5c}Lawniczak, A.T. \& Tang, X.W., Packet Traffic Dynamics Near Onset of Congestion in Data Communication Network Model, \emph{Acta Physica Polonica B}, \textbf{37}:5 (2006) 1579:1604.
\bibitem{DFA1}Lee, C.Y., Higher-order correlations in data network traffic, \emph{J. Korean Phys. Soc.}, \textbf{45}:6 (2004) 1664-1670.
\bibitem{selfsimilar1}Leland, W.E., \& Wilson, D.V., High time-resolution measurement and analysis of LAN traffic: Implications for LAN interconnection, \emph{Proceedings IEEE lNFOCOM '91} (Bal Harbour, Florida, 1991) pp. 1360-66.
\bibitem{selfsimilar2}Leland, W.E., Taqqu, M.S., Willinger, W., \& Wilson, D.V., On the self-similar nature of Ethernet traffic (extended version) \emph{IEEE/ACM Transactions on Networking}, \textbf{2}:1 (1994)  1-15.
\bibitem{spectrum14}Li, L. \& Lee, G., DDoS Attack Detection and Wavelets, \emph{Telecom. Sys.}, \textbf{28}:4 (2005) 435-451.
\bibitem{netcriticnew1}Li, L., Alderson, D.L., Willinger, W., \&
Doyle, J., A first principles approach to understanding the
Internet's router-level topology, \emph{ACM SIGCOMM ComComm. Rev},
\textbf{34}:4, 3-14 (2004).
\bibitem{netcriticnew2}Li, L., DL Alderson, Willinger, W., \&
Doyle, J., Towards a theory of scale-free graphs: Definition,
properties, and implications, \emph{Internet Mathematics},
\textbf{2}:4, 431-523 (2006).
\bibitem{TCPred3}Low, S.H., Paganini, F., Wang, J.T., \& Doyle, J.C., Linear stability of TCP/RED and a scalable control, \emph{Computer Networks}, \textbf{43}:5 (2003) 633-647.
\bibitem{netcriticnew6}Mahadevan, P., Krioukov, D., Fomenkov, M., Huffaker, B., Dimitropolous, X., Claffy, K.C., \& Vahdat, A.,
Lessons from three views of the Internet topology preprint
arXiv:cs/0508033 (2005).
\bibitem{selfsimilarcause2b}Mandelbrot, B.B., Long-run linearity, locally Gaussian processes, H-spectra and infinite variances, \emph{Intl. Econ. Rev.}, \textbf{10} (1969) 82-111.
\bibitem{TCPthrough3}Mathis, M., Semke, J., Madhavi, J., \& Ott, T., The Macroscopic Behavior of the TCP Congestion Avoidance Algorithm, \emph{ACM SIGCOMM ComComm. Rev.}, \textbf{27}:3 (1997) 67-82.
\bibitem{comp1}Maurer, S.M. \& Huberman, B.A., Competitive dynamics of web sites, \emph{Journal of Economic Dynamics and Control}, \textbf{27}:12 (2003) 2195-2206.
\bibitem{mawi}MAWI, Widely Integrated Distributed Environment (WIDE) Project, Kanagawa, Japan, MAWI Working Group Traffic Archive (WIDE) Backbone traffic traces) http://mawi.wide.ad.jp/mawi/
\bibitem{fluctuations5}Meloni, S., Gom\'{e}z-Garde\~{n}as, J., Latora, V., \& Moreno, Y., Scaling Breakdown in Flow Fluctuations on Complex Networks, \emph{Phys. Rev. Lett.}, \textbf{100} (2008) 208701
\bibitem{TCPred4}Misra, V., Gong, W.B., \& Towsley, D., Fluid-based analysis of a network of AQM routers supporting TCP flows with an application to RED, \emph{ACM/SIGCOMM ComComm Review}, \textbf{30}:4 (2000) 151-160.
\bibitem{powerlawsense2}Mitzenmacher, M., A brief history of generative models for power law and lognormal distributions, \emph{Internet Mathematics}, \textbf{1}:2, 226-251 (2004).
\bibitem{selfsimilarcause5c}Mitzenmacher, M., Dynamic models for file sizes and double Pareto distributions,
\emph{Internet Mathematics}, \textbf{1}:3, 305-333 (2004).
\bibitem{powerlawsense3}Mitzenmacher, M., Editorial: The future of power law research
, \emph{Internet Mathematics}, \textbf{2}:4, 525-534 (2005).
\bibitem{ID4}Moore, A. \& Papagiannaki, K., Toward the accurate identification of network applications, in \emph{Passive and Active Network Measurement} edited by Dovrolis, C. (Springer, Berlin, 2005) pp. 41-54.
\bibitem{slammer}Moore, D., Paxson, V., Savage, S., Shannon, C., Staniford, S., \& Weaver, N., Inside the Slammer Worm, \emph{IEEE Security \& Privacy}, \textbf{1}:4  (2003) 33-39.
\bibitem{epidemic4}Moore,D., Shannon, C., Voelker, G., \& Savage, S., Internet Quarantine: Requirements for Containing Self-Propagating Code, in \emph{Proceedings of INFOCOM 2003}, \textbf{3} (San Francisco ,2003) pp.1901-1910.
\bibitem{cascade2}Moreno, Y., Gomez, J.B., \& Pacheco, A.F., Instability of scale-free networks under node-breaking avalanches, \emph{Europhys. Lett.}, \textbf{58} (2002)  630-636.
\bibitem{netcongest6d}Moreno, Y., Pastor-Satorras, R., V\'{a}quez, A., \& Vespignani, A., Critical load and congestion instabilities in scale-free networks, \emph{Europhys. Lett.}, \textbf{62} (2003) 292-298.
\bibitem{elephant4}Mori, T., Kawahara, R., Naito, S., \& Goto, S., On the Characteristics of Internet Traffic Variability: Spikes and Elephants, \emph{Proceedings of the 2004 International Symposium on Applications and the Internet} (Tokyo, Japan, 2004) pp. 99-106.
\bibitem{infonet}Moses, M.E., Forrest, S., Davis, A.L., Lodder, M.A., \& Brown, J.H., Scaling theory for information networks, \emph{J. Royal Soc. Interface}, 5:29, (2008) 1469-1480.
\bibitem{cascade1}Motter, A.E. \& Lai, Y.C., Cascade-based attacks on complex networks, \emph{Phys. Rev. E}, \textbf{66} (2002) 065102.
\bibitem{spectrum2e}Mukherjee, A., On The Dynamics and Significance of Low Frequency Components of Internet Load, University of Pennsylvania Technical Reports, MS-CIS-92-83 (1992).
\bibitem{netcongest4c}Mukherjee, G. \& Manna, S., Phase transition in a directed traffic flow network, \emph{Phys. Rev. E}, \textbf{71} (2005) 066108.
\bibitem{newman1}Newman, M.E.J., Scientific collaboration networks.  I. Network construction and fundamental results, \emph{Phys. Rev. E}, \textbf{64} (2001) 016131.
\bibitem{review2}Newman, M.E.J., The structure and function of complex networks, \emph{SIAM Review}, \textbf{45} (2003) 167-256.
\bibitem{wavelet2}Nievergelt, Y., \emph{Wavelets Made Easy}, (Springer, Berlin, 1999).
\bibitem{netcongest4}Ohira, T. \& Sawatari, R., Phase transition in a computer network traffic model, \emph{Phys. Rev. E}, \textbf{58} (1998) 193-195.
\bibitem{spectrum2f}Owezarski, P., \& Larrieu, N., Internet Traffic Characterization - An Analysis of Traffic Oscillations, in \emph{High Speed Networks and Multimedia Communications} edited by MM Freire, P Lorenz, \& M Lee, (Springer, Berlin, 2004) pp. 96.
\bibitem{TCPthrough4}Padhye, J., Firoiu, V., Towsley, D., \& Kurose, J., Modeling TCP Reno performance: a simple model and its empirical validation, \emph{IEEE/ACM Transactions on Networks}, \textbf{8}:2 (2000) 133-145.
\bibitem{elephant5}Papagiannaki, K., Taft, N., Bhattacharyya, S., Thiran, P., Salamatian, K. \& Diot, C., A pragmatic definition of elephants in internet backbone traffic, \emph{Proceedings of the 2nd ACM SIGCOMM Workshop on Internet Measurement} (Marseille, France, 2002) pp. 175-176.
\bibitem{spectrum2c}Papagiannaki, K., Taft, N., Zhang, Z., \& Diot, C., Long-term forecasting of Internet backbone traffic, \emph{IEEE Transactions on Neural Networks}, \textbf{16}:5 (2005) 1110-1124.
\bibitem{selfsimilar5} Park, K. \& Willinger, W., Self-similar network traffic: an overview, in \emph{Self-Similar Network Traffic and Performance Evaluation}, edited by Park, K. \& Willinger, W. (John Wiley \& Sons, New York, 2000) pp. 1.
\bibitem{selfsimilarcause3c}Park, K., Kim, G., \& Crovella, M.E., The protocol stack and its modulating effect on self-similar traffic, in \emph{Self-Similar Network Traffic and Performance Evaluation}, edited by Park, K. \& Willinger, W., (John Wiley \& Sons, New York, 2000) pp. 349.
\bibitem{selfsimilarcause3}Park, K., Kim, G., \& Crovella, M.E., On the relationship between file sizes, transport protocols, and self-similar network traffic, in \emph{Proceedings of the Fourth International Conference on Network Protocols (ICNP'96)}  (Columbus, OH, 1996) pp. 171-180.
\bibitem{IBM TCP}Parziale, L., Britt, D.T., Davis, C., Forrester, J., Liu, W., Matthews, C., \& Rosselot, N., \emph{TCP/IP Tutorial and Technical Overview}, IBM: Internal Technical and Support Organization (2006).
\bibitem{vespignani1}Pastor-Satorras, R. \& Vespignani, A., Epidemic spreading in scale-free networks, \emph{Phys. Rev. Lett.}, \textbf{86} (2001) 3200-3203.
\bibitem{review7}Pastor-Satorras, R. \& Vespignani, A., \emph{Evolution and Structure of the Internet: A Statistical Physics Approach}, (Cambridge University Press, New York, 2004).
\bibitem{protocol4}Paxson, V., Growth trends in wide-area TCP connections,  \emph{IEEE Network}, \textbf{8}:4 (1994) 8-17.
\bibitem{netcriticnew4}Paxson, V., Strategies for sound Internet
measurement in Proceedings of the 4th ACM SIGCOMM conference on
Internet measurement, (Sicily, Italy, 2004) pp. 263-271.
\bibitem{selfsimilar3}Paxson, V., \& Floyd, S., Wide area traffic: the failure of Poisson modeling, \emph{IEEE/ACM Transactions on Networking}, \textbf{3}:3 (1995) 226-244.
\bibitem{wavelet1}Percival, D.B. \& Walden, A.T., \emph{Wavelet Methods for Time Series Analysis}, (Cambridge University Press, New York, 2000).
\bibitem{TCPred2}Ranjan, P., Abed, E.H., \& La, R.J., Nonlinear instabilities in TCP-RED, \emph{IEEE/ACM Transactions on Networking}, \textbf{12}:6 (2004) 1079-1092.
\bibitem{PingER}Stanford SLAC PingER (Ping end-to-end reporting) Project www-iepm.slac.stanford.edu/pinger/
\bibitem{multifractal0}Riedi, R. \& L\'{e}vy-V\'{e}hel, J., TCP traffic is multifractal: A numerical study, preprint (1997).
\bibitem{selfsimilar9}Roberts, J.W., Traffic theory and the Internet, \emph{IEEE Communications Magazine}, \textbf{39}:1 (2001)  94-99.
\bibitem{spectrum2d}Roughan, M., Greenberg, A., Kalmanek, C., Rumsewicz, M., Yates, J., \& Zhang, Y., Experience in measuring backbone traffic variability: models, metrics, measurements and meaning, in \emph{Proceedings of the 2nd ACM SIGCOMM Workshop on Internet Measurement} (Marseille, France, 2002) pp. 91-92.
\bibitem{bgpstorm2}Roughan, M., Li, J., Bush, R., Mao, Z.Q., \& Griffin, T., Is BGP Update Storm a Sign of Trouble: Observing the Internet Control and Data Planes During Internet Worms, in \emph{Proceedings of IEEE SPECTS'06} (Calgary, Alberta, 2006).
\bibitem{spectrum2}Shoch, J.F., \& Hupp, J.A., Measured performance of an Ethernet local network, \emph{Communications of the ACM}, \textbf{23}:12 (1980) 711-721.
\bibitem{review6}Shan, X.M., Wang, L., Ren, Y., Yuan, J., \& Song, Y.H., Advances in the research of Internet complexity,  \emph{Journal of Beijing University of Posts and Telecommunications}, \textbf{29}:1 (2006) 1-8 (in Chinese).
\bibitem{book2}Shirow, M., \emph{The Ghost In The Shell (Kokaku Kidotai)}, (Kodansha, Tokyo, 1991).
\bibitem{TCPchaos7}Sikdar, B. \& Vastola, K., The effect of TCP on the self-similarity of network traffic, in \emph{Proceedings of the Conference of Information Science Systems}, John Hopkins University, (Baltimore, MD, 2001).
\bibitem{TCPchaos6} Sikdar, B. \& Vastola, K., On the contribution of TCP to the self-similarity of network traffic in \emph{Evolutionary Trends of the Internet}, (Springer, Berlin, 2001) pp. 596.
\bibitem{transcritical}Smith, R.D., Data traffic dynamics and saturation on a single link, \emph{International Journal of Computer Systems Science \& Engineering}, {3}:1, 11-16 (2009).
\bibitem{netcongest6}Sol\'{e}, R. \& Valverde, S., Information transfer and phase transitions in a model of internet traffic, \emph{Physica A}, \textbf{289}:4 (2001) 595-605.
\bibitem{netcongest6b}Sol\'{e}, R. \& Valverde, S., Self-organized critical traffic in parallel computer networks, \emph{Physica A}, \textbf{312}:4 (2002) 636-648.
\bibitem{review5}Strogatz, S.H., Exploring complex networks, \emph{Nature}, \textbf{410} (2001) 268-276.
\bibitem{DFA2}Tadaki, S., Long-Term Power-Law Fluctuation in Internet Traffic,  J. Phys. Soc. Japan, \textbf{76} (2007) 044001.
\bibitem{netroute4c}Tadi\'{c}, B. \& Thurner, S., Information super-diffusion on structured networks, \emph{Physica A}, \textbf{332} (2004) 566-584.
\bibitem{netroute4d} Tadi\'{c}, B. \& Thurner, S., Search and topology aspects in transport on scale-free networks, \emph{Physica A}, \textbf{346}:2 (2005) 183-190.
\bibitem{netroute4b}Tadi\'{c}, B., Thurner, S., \& Rodgers, G.J., Traffic on complex networks: Towards understanding global statistical properties from microscopic density fluctuations, \emph{Phys. Rev. E}, \textbf{69}
(2004) 036102.
\bibitem{netroute4b2}Tadi\'{c}, B., Thurner, S., \& Rodgers, G.J., Transport on complex networks: flow, jamming and optimization, \emph{International Journal of Bifurcation and Chaos}, \textbf{17}:7 (2007) 2363-2385.
\bibitem{netcongest2b}Takayasu, M., Takayasu, H., \& Fukuda, K., Origin of critical behavior in Ethernet traffic, \emph{Physica A}, \textbf{287}:2 (2000) 289-301.
\bibitem{netcongest3}Takayasu, M., Takayasu, H., \& Fukuda, K., Dynamic phase transition observed in the Internet traffic flow, \emph{Physica A}, \textbf{277}:2 (2000) 248-255.
\bibitem{netcongest2}Takayasu, M., Takayasu, H.,  \& Sato, T., Critical behaviors and 1/f noise in information traffic, \emph{Physica A}, \textbf{233}:3 (1996) 824-834.
\bibitem{netcongest2a}Takayasu, M., Takayasu, H., \& Tretyakov, A.Y., Phase transition in a computer network model, \emph{Physica A}, \textbf{253}:1 (1998) 315-322.
\bibitem{multifractal1}Taqqu, M.S., Teverovsky, V., \& Willinger, W., Is network trac self-similar or multifractal,  \emph{Fractals}, \textbf{5} (1997) 63-73.
\bibitem{packetsize2}Thompson, K., Miller, G.J., \& Wilder, R., Wide-area Internet traffic patterns and characteristics,  \emph{IEEE Network}, \textbf{11}:6 (1997) 10-23.
\bibitem{multifractal5}Uhlig, S., Non-stationarity and high-order scaling in TCP flow arrivals: a methodological analysis, \emph{ACM Comm. Rev.}, \textbf{34}:2 (2004) 9-24.
\bibitem{multifractal6}Veitch, D., Hohn, N., \& Abry, P., Multifractality in TCP/IP traffic: the case against, \emph{Computer Networks}, \textbf{48}:3 (2005) 293-313.
\bibitem{TCPchaos1}Veres, A. \& Boda, M., The chaotic nature of TCP congestion control in the \emph{Proceedings of IEEE INFOCOM 2000}, \textbf{3} (Tel-Aviv, Brazil, 2000) pp. 1715-1723.
\bibitem{TCPchaos10}Veres, A., Kenesi, Z., Moln\'{a}r, S., \& Vattay, G., On the propagation of long-range dependence in the Internet, in \emph{ACM SIGCOMM ComComm. Rev.}, \textbf{30}:4 (2000) 243-254.
\bibitem{singularity}Vinge, V., Signs of the Singularity, \emph{IEEE Spectrum}, \textbf{45}:6 (2008) 76-82.
\bibitem{comp4}Wang, Y.S. \& Wu, H., \emph{Physica A}, Dynamics of a cooperation-competition model for the WWW market, \textbf{339}:4 (2004) 609-620.
\bibitem{mutualism}Wang, Y.S. \& Wu, H., \emph{Physica A}, Dynamics of a macroscopic model characterizing mutualism of search engines and web sites, \textbf{363}:2 (2006) 537-550.
\bibitem{netroute5also}Wang, W.X., Yin, C.Y., Yan, G., \& Wang, B.H., Integrating local static and dynamic information for routing traffic, \emph{Phys. Rev. E}, \textbf{74} (2006) 016101.
\bibitem{epidemic5}Wang, Y., Chakrabarti, D., Wang, C.X., Faloutsos, C., Epidemic Spreading in Real Networks: An Eigenvalue Viewpoint, in \emph{Proceedings of the 22nd International Symposium on Reliable Distributed Systems (SRDS'03)} (Florence, 2003) pp. 25-44.
\bibitem{watts1}Watts, D.J. \& Strogatz, S.H., Collective dynamics of `small-world' networks, \emph{Nature} \textbf{393} (1998) 440442
\bibitem{leaves}Whitman, W., When I heard the Learn'd Astronomer, \emph{Leaves of Grass}, (Bantam, New York, 1983).
\bibitem{selfsimilarcause1}Willinger, W., Taqqu, M.S., Sherman, R., \& Wilson, D.V., Self-similarity through high-variability: statistical analysis of Ethernet LAN traffic at the source level, \emph{ACM SIGCOMM ComComm. Rev.}, \textbf{25}:4 (1995) 100-113.
\bibitem{selfsimilarcause2} Willinger, W., Taqqu, M.S., Sherman, R., \& Wilson, D.V., Self-Similarity through High-Variability: Statistical Analysis of Ethernet LAN Traffic at the Source Level, \emph{IEEE/ACM Transactions on Networking}, \textbf{5}:1 (1997) 71-86.
\bibitem{selfsimilar11}Willinger, W., Paxson, V., \& Taqqu, M.S., Self-similarity and heavy tails: Structural modeling of network traffic, in \emph{A Practical Guide To Heavy Tails: Statistical Techniques and Applications} edited by RF Adler, RE. Feldman, \& MS. Taqqu, (Birkhauser, Boston, 1998), pp. 27-53.
\bibitem{selfsimilar7}Willinger, W., Govindan, R., Jamin, S., Paxson, V., \& Shenker, S., Scaling phenomena in the Internet: Critically examining criticality,  \emph{Proc. Natl. Acad. of Sci.}, \textbf{99} (2002) 2573-2580.
\bibitem{bookcritic}Willinger, W., Paxson, V., Riedi, R., \& Taqqu, M.S., Long-range dependence and data network
traffic in Theory and Applications of Long-range Dependence edited
by Doukhan, P., Oppenheim, G., \& Taqqu, M.S. (Birkhauser, Boston,
2003) pp. 373-408.
\bibitem{traffichistory}Willinger, W., \& Paxson, V., Where Mathematics Meets the Internet, \emph{Notices of the AMS}, \textbf{45}:8 (1998) 961-970.
\bibitem{netcriticAMS}Willinger, W., Alderson, D., \& Doyle, J.C., ``Mathematics and the Internet: A Source of Enormous Confusion and Great Potential'', \emph{Notices of the AMS}, \textbf{56}:5, p. 586-599.
\bibitem{netcongest6b2}Woolf, M., Arrowsmith, D.K., Mondrag\'{o}n, R.J., \& Pitts, J.M., Optimization and phase transitions in a chaotic model of data traffic, \emph{Phys. Rev. E}, \textbf{66} (2002) 046106.
\bibitem{epidemic2}Yan, G. et. al., Epidemic spread in weighted scale-free networks, \emph{Chinese Physics Letters}, \textbf{22}:2, (2005) 510-513. 
\bibitem{netroute3}Yan, G., Zhou, T., Hu, B., Fu, Z.Q., \& Wang, B.H., Efficient routing on complex networks, \emph{Phys. Rev. E}, \textbf{73} (2006) 046108.
\bibitem{netroute5}Yin, C.Y., Wang, B.H., Wang, W.X., Yan, G., \& Yang, H.J., Traffic dynamics based on an efficient routing strategy on scale free networks, \emph{Eur. Phys. J. B}, \textbf{49}:2 (2006) 205-211.
\bibitem{netcongest7}Yuan, J. \& Mills, K.J., Exploring Collective Dynamics in Communication Networks, \emph{J. Res. Natl. Inst. Stand. Technol.}, \textbf{107}:2 (2002) 179-191.
\bibitem{netcongest8} Yuan, J. \& Mills, K.J., A cross-correlation-based method for spatial–temporal traffic analysis, \emph{Performance Evaluation}, \textbf{61}:2 (2005) 163-180.
\bibitem{netcongest9} Yuan, J. \& Mills, K.J., Macroscopic Dynamics in Large-Scale Data Networks, in  \emph{Complex Dynamics in Communication Networks}, (Springer, Berlin, 2005) pp. 191-211.
\bibitem{ddos}Yuan, J. \& Mills, K.J., Monitoring the macroscopic effect of DDoS flooding attacks, \emph{IEEE Transactions on Dependable and Secure Computing}, \textbf{2}:4 (2005) 324-355.
\bibitem{netcongest10} Yuan, J. \& Mills, K.J., Simulating Timescale Dynamics of Network Traffic Using Homogeneous Modeling, \emph{J. Res. Natl. Inst. Stand. Technol.}, \textbf{111}:3 (2006) 227-242.
\bibitem{netcongest12}Yuan, J., Yong, R., \& Shan, X.M., Analysis of a type of computer network cellular automata model, \emph{Acta Physica Sinica}, \textbf{49}:3 (2000) 398-402 (in Chinese).
\bibitem{netcongest5b}Yuan, J., Ren, Y., Liu, F., \& Shan, X.M., Phase transition and collective correlation behavior in the complex computer network, \emph{Acta Physica Sinica}, \textbf{50}:7 (2001) , 1221-1225 (in Chinese).
\bibitem{netcongest11}Yuan, J., Wang, J., Xu, Z.X., \& Li, B., Time-dependent collective behavior in a computer network model, \emph{Physica A}, \textbf{368}:1 (2006) 294-304.
\bibitem{netcongest6c}Zhao, L., Lai, Y.C., Park, K.H., \& Ye, N., Onset of traffic congestion in complex networks, \emph{Phys. Rev. E}, \textbf{71} (2005) 026125.
\bibitem{netroute7}Zhu, D.P., Gritter, M., \& Cheriton, D.R., Feedback based routing, \emph{ACM SIGCOMM ComComm. Rev.}, \textbf{33}:1 (2003) 71-76.
\bibitem{DFA3}Zhu, X.Y., Liu, Z.H., \& Tang, M., Detrended Fluctuation Analysis of Traffic Data, \emph{Chin. Phys. Lett.}, \textbf{24} (2007) 2142-2145.
\end{thebibliography}
\end{document}